\documentclass[reprint,nocopyrightspace]{sigplanconf}

\usepackage{amsmath, amsthm, amssymb}
\usepackage{stmaryrd}
\usepackage{bussproofs}

\newcommand{\ignore}[1]{}
\newcommand{\ital}[1]{{\em #1}}

\newcommand{\eg}{{\rm e.g.},}
\newcommand{\ra}{\rightarrow}

\newcommand{\lprolog}{$\lambda$Prolog}
\newcommand{\hhf}{$hohh$}
\newcommand{\lftype}{\text{\ital{lf-type}}}
\newcommand{\lfobject}{\text{\ital{lf-obj}}}

\newcommand{\encInv}[2]{inv_{#2}(#1)}
\newcommand{\invprove}[5]{#1;#3 \longrightarrow_{#2} \oftype{#4}{#5}}
\newcommand{\invabs}{\mbox{\sl inv-abs}}
\newcommand{\invconst}{\mbox{\sl inv-const}}
\newcommand{\invvar}{\mbox{\sl inv-var}}

\newcommand{\ruvappt}{{\scriptsize APP$_\text{t}$}}
\newcommand{\ruvpit}{{\scriptsize PI$_\text{t}$}}
\newcommand{\ruvctxt}{{\scriptsize CTX$_\text{t}$}}
\newcommand{\ruvinito}{{\scriptsize INIT$_\text{o}$}}
\newcommand{\ruvappo}{{\scriptsize APP$_\text{o}$}}
\newcommand{\ruvabso}{{\scriptsize ABS$_\text{o}$}}

\renewcommand{\vec}[1]{\overrightarrow{#1}}

\newcommand{\subst}[2]{#1[#2]}

\newcommand{\iprove}[2]{\sequent{#1}{#2}}
\newcommand{\lfprove}[2]{#1 \, \vdash \, #2}
\newcommand{\hastype}[2]{hastype \ #1 \ #2}

\newcommand{\encEq}[2]{#1 \sim #2}

\newcommand{\encTerm}[1]{\langle #1 \rangle}
\newcommand{\enc}[1]{\lbrace\!\!\lbrace #1 \rbrace\!\!\rbrace}
\newcommand{\encExtP}[2]{\llbracket #1 \rrbracket^{+}_{#2}}
\newcommand{\encExtN}[1]{\llbracket #1 \rrbracket^{-}}
\newcommand{\termUV}[2]{#1 \sqsubset_o #2}
\newcommand{\formulaUV}[2]{#1 \sqsubset_t #2}

\newcommand{\forallx}[2]{\forall #1.#2}
\newcommand{\sequent}[2]{#1 \longrightarrow #2}
\newcommand{\sigsequent}[3]{#1 ; #2 \longrightarrow #3}
\newcommand{\focusedsigseq}[4]{#1 ; #2 \buildrel #4 \over
  \longrightarrow #3}
\newcommand{\lambdax}[2]{\lambda #1.#2}
\newcommand{\typedlambda}[3]{\lambda #1 \mbox{:} #2 . #3}

\newcommand{\typedpi}[3]{\Pi #1 \mbox{:} #2 . #3}
\newcommand{\typedpis}[3]{\Pi \vec{#1 \mbox{:} #2} . #3}
\newcommand{\app}[2]{#1\ #2}
\newcommand{\emptyctx}{\cdot}
\newcommand{\oftype}[2]{#1 : #2}
\newcommand{\ctx}{\mbox{\sl ctx}}
\newcommand{\sig}{\mbox{\sl sig}}
\newcommand{\kind}{\mbox{\sl kind}}
\newcommand{\type}{\mbox{\sl Type}}

\newcommand{\bcGoal}{\mbox{\sl backchain}}
\newcommand{\topGoal}{$\top \mbox{\sl R}$}

\newcommand{\impGoal}{$\supset\! \mbox{\sl R}$}
\newcommand{\allGoal}{$\forall \mbox{\sl R}$}
\newcommand{\atomicGoal}{$\mbox{\sl decide}$}
\newcommand{\initial}{$\mbox{\sl init}$}
\newcommand{\allClause}{$\forall \mbox{\sl L}$}
\newcommand{\impClause}{$\supset\! \mbox{\sl L}$}

\newcommand{\nullsig}{\mbox{\sl null-sig}}
\newcommand{\kindsig}{\mbox{\sl kind-sig}}
\newcommand{\typesig}{\mbox{\sl type-sig}}
\newcommand{\nullctx}{\mbox{\sl null-ctx}}
\newcommand{\typectx}{\mbox{\sl type-ctx}}
\newcommand{\typekind}{\mbox{\sl type-kind}}
\newcommand{\pikind}{\mbox{\sl pi-kind}}
\newcommand{\varfam}{\mbox{\sl var-fam}}
\newcommand{\varobj}{\mbox{\sl var-obj}}
\newcommand{\pifam}{\mbox{\sl pi-fam}}
\newcommand{\absfam}{\mbox{\sl abs-fam}}
\newcommand{\appfam}{\mbox{\sl app-fam}}
\newcommand{\appobj}{\mbox{\sl app-obj}}
\newcommand{\absobj}{\mbox{\sl abs-obj}}

\newcommand{\ie}{{\rm i.e.}}
\newcommand{\etal}{{\rm et. al.\ }}

\newtheorem{theorem'}[theorem]{Theorem}
\newtheorem{corollary'}{Corollary}
\newtheorem{lemma'}[theorem]{Lemma}

\newtheorem{claim}{Claim}

\title{Translating Specifications in a Dependently Typed\\
 Lambda Calculus into a Predicate Logic Form}
\authorinfo{Mary Southern \and Gopalan Nadathur}
           {Computer Science and Engineering\\
            University of Minnesota\\
            Minneapolis, MN 55455}
           {}

\begin{document}
\maketitle

  \begin{abstract}
  Dependently typed lambda calculi such as the Edinburgh Logical
Framework (LF) are a popular means for encoding rule-based
specifications concerning formal syntactic objects.
In these frameworks, relations over terms representing formal objects
are naturally captured by making use of the dependent structure of types.
We consider here the meaning-preserving translation of specifications
written in this style into  a predicate logic over simply typed
$\lambda$-terms. 
Such a translation can provide the basis for
efficient implementation and sophisticated capabilities for reasoning
about specifications.
We start with a previously described translation of LF specifications
to formulas in the logic of higher-order hereditary Harrop (\hhf) formulas.
We show how this translation can be improved by recognizing
and eliminating redundant type checking information contained in it.
This benefits both the clarity of translated formulas, and reduces the
effort which must be spent on type checking during execution.
To allow this translation to be used to execute LF specifications,
we describe an inverse transformation from \hhf-terms to LF
expressions; thus computations can be carried out using the translated
form and the results can then be exported back into LF.
Execution based on LF specifications may also involve some forms of type
reconstruction.
We discuss the possibility of supporting such a capability using the
translation under some reasonable restrictions on the structure of
specifications.

  \end{abstract}

  \section{Introduction}
\label{sec:intro}

The Edinburgh Logical Framework (LF) has proven to be a useful device
for specifying formal systems such as logics and programming
languages.
At its core, LF is a dependently typed lambda calculus.
By exploiting the abstraction operator that is part of the syntax of
LF, it is possible to succinctly encode formal objects whose structure
embodies binding notions.
Because types can be indexed by terms, we can use them to express
relations between the formal objects encoded in terms.
If we view such types as formulas, terms that have a given type can be
interpreted as proofs of the formula that type represents.
Thus, LF specifications can be given a logic programming
interpretation using a notion of proof search that corresponds to
determining inhabitation of given types.
The Twelf system is an implementation of LF that is based on such an
interpretation.

An alternative approach to specifying formal systems is to use a
predicate logic.
Objects treated by the formal systems can be represented by the terms
of this logic and relations between them can be expressed through
predicates over these terms.
If the terms include a notion of abstraction (\eg~if they encompass
simply typed lambda terms) they provide a convenient means for
representing binding notions.
While an unrestricted predicate logic would be capable of describing
relations adequately, it is preferable to limit the permitted
formulas so that the desired interpretation of rule based
specifications can be modeled via a constrained proof search
behavior.
The logic of higher-order hereditary Harrop formulas (\hhf) has been
designed with these ideas in mind and many experiments have shown this
logic to be a useful specification device (\eg~see
\cite{miller12proghol}).
This logic has also been given a computational interpretation in the
language \lprolog~\cite{nadathur88iclp}.
Moreover, an efficient implementation of \lprolog~ has been developed
in the Teyjus system \cite{teyjus.website}.

There are obvious similarities between the two different
approaches to specification, making it interesting to explore the
connections between them more formally.
In early work, Felty~\etal showed that LF derivations could be encoded
in \hhf~derivations by describing a translation from the former to the
latter \cite{felty90cade}.
This translation demonstrated the expressive power of \hhf, but was
not directly usable in relating proof search behavior.
To rectify this situation, Snow~\etal showed how to translate LF
specifications into \hhf~formulas in such a way that the process of
constructing a derivation could be related \cite{snow10ppdp}.
This work provided the basis for an alternative implementation of
Twelf.
The translation also has the potential to be useful in bringing the
power of the Abella prover \cite{abella.website} to bear on reasoning
about Twelf specifications.

This paper continues the work described in \cite{snow10ppdp}. There
are four specific contributions it makes in this setting:
\begin{enumerate}
\item An important part of the translation is the recognition and
  elimination of redundant typing information in specifications. We
  describe an improvement to the criterion presented in
  \cite{snow10ppdp} for this purpose.

\item In contrast to~\cite{snow10ppdp}, we show how to modularize the
  proof of redundancy of typing information, establishing a result
  concerning LF first and then lifting this result to the
  translation. This enables us to present results that also apply
  directly to LF.

\item If we are to use the translation as a means for implementing
  proof search in Twelf, we need also a way to return to Twelf
  expressions after completing execution in \lprolog. We
  describe such an inverse transformation.

\item Logic programming in Twelf includes a process of type
  reconstruction. We begin an analysis of the translation towards
  understanding whether type reconstruction on the translated
  expression will agree with Twelf's behavior. This analysis is
  incomplete, but we believe the approach to be sound and the
  remaining work to be mainly that of elaborating the details.
\end{enumerate}

The next two sections describe LF and the \hhf~logic respectively and
discuss their computational interpretations.
Section~\ref{sec:naive} then presents a simple translation of LF
specifications into \hhf~ ones.
The following section takes up the task of improving this
translation.
In particular, it characterizes certain bound variable occurrences in
types using a notion called {\em strictness} and uses this
characterization to identify redundancy in typing.
We are then able to eliminate such redundancy in translation.
Section~\ref{sec:inverse} describes the inverse translation from
\hhf~terms found via proof search to LF expressions in the originating
context for the translation.
Section~\ref{sec:existential} contains a discussion on the treatment
of type reconstruction in Twelf proof search.
We end the paper with a discussion of future directions to this work
in Section~\ref{sec:conclusion}.

  \section{Logical Framework}
\label{sec:lf}
\begin{figure*}
\begin{center}
\begin{tabular}{ccc}
   \AxiomC{}
   \RightLabel{\nullsig}
   \UnaryInfC{$\cdot~sig$}
   \DisplayProof

&

   \AxiomC{$\Sigma~sig \quad a \not\in dom(\Sigma) \quad \lfprove{}{A~\kind}$}
   \RightLabel{\kindsig}
   \UnaryInfC{$\Sigma,a:K~sig$}
   \DisplayProof

&

   \AxiomC{$\Sigma~sig \quad c\not\in dom(\Sigma) \quad \lfprove{}{A:\type}$}
   \RightLabel{\typesig}
   \UnaryInfC{$\Sigma,c:A~sig$}
   \DisplayProof
\end{tabular}

\vspace{10pt}

\begin{tabular}{cc}
   \AxiomC{$\Sigma~sig$}
   \RightLabel{\nullctx}
   \UnaryInfC{$\lfprove{}{\emptyctx\ \ctx}$}
   \DisplayProof

&

   \AxiomC{$\lfprove{\Gamma}{\oftype{A}{\type}} \quad
     \lfprove{}{\Gamma\ \ctx}\quad x \notin dom(\Gamma)$}
   \RightLabel{\typectx}
   \UnaryInfC{$\lfprove{}{\Gamma, x : A\ \ctx}$}
   \DisplayProof
\end{tabular}

\vspace{10pt}

\begin{tabular}{cc}
   \AxiomC{$\lfprove{}{\Gamma\ \ctx}$}
   \RightLabel{\typekind}
   \UnaryInfC{$\lfprove{\Gamma}{\type\ \kind}$}
   \DisplayProof

&
   \AxiomC{$\lfprove{\Gamma}{\oftype{A}{\type}} \quad
     \lfprove{\Gamma,\oftype{x}{A}}{K\ \kind}$}
   \RightLabel{\pikind}
   \UnaryInfC{$\lfprove{\Gamma}{\typedpi{x}{A}{K}\ \kind}$}
   \DisplayProof
\end{tabular}

\vspace{10pt}

\begin{tabular}{ccc}
   \AxiomC{$\lfprove{}{\Gamma\ \ctx} \quad \oftype{u}{K} \in \Gamma$}
   \RightLabel{\varfam}
   \UnaryInfC{$\lfprove{\Gamma}{\oftype{u}{K^\beta}}$}
   \DisplayProof

&
   \AxiomC{$\lfprove{\Gamma}{\oftype{A}{\type}} \quad \lfprove{\Gamma,
       \oftype{x}{A}}{\oftype{B}{\type}}$}
   \RightLabel{\pifam}
   \UnaryInfC{$\lfprove{\Gamma}{\oftype{(\typedpi{x}{A}{B})}{\type}}$}
   \DisplayProof

&
   \AxiomC{$\lfprove{\Gamma}{\oftype{A}{\typedpi{x}{B}{K}}} \quad
     \lfprove{\Gamma}{\oftype{M}{B}}$}
   \RightLabel{\appfam}
   \UnaryInfC{$\lfprove{\Gamma}{\oftype{(\app{A}{M})}{(K[M/x])^\beta}}$}
   \DisplayProof
\end{tabular}

\vspace{10pt}

\begin{tabular}{ccc}
   \AxiomC{$\lfprove{}{\Gamma\ \ctx} \quad \oftype{x}{A} \in \Gamma$}
   \RightLabel{\varobj}
   \UnaryInfC{$\lfprove{\Gamma}{\oftype{x}{A^\beta}}$}
   \DisplayProof

&
   \AxiomC{$\lfprove{\Gamma}{\oftype{A}{\type}} \quad
     \lfprove{\Gamma,\oftype{x}{A}}{\oftype{M}{B}}$}
   \RightLabel{\absobj}
   \UnaryInfC{$\lfprove{\Gamma}{\oftype{(\typedlambda{x}{A}{M})}{(\typedpi{x}{A^\beta}{B})}}$}
   \DisplayProof

&

   \AxiomC{$\lfprove{\Gamma}{\oftype{M}{\typedpi{x}{A}{B}}} \quad
     \lfprove{\Gamma}{\oftype{N}{A}}$}
   \RightLabel{\appobj}
   \UnaryInfC{$\lfprove{\Gamma}{\oftype{(\app{M}{N})}{(B[N/x])^\beta}}$}
   \DisplayProof
\end{tabular}
\end{center}
\nocaptionrule  \caption{Rules for Inferring LF Assertions}
  \label{fig:lf-rules}
\end{figure*}
%
This section introduces dependently typed $\lambda$-calculi as a means
for specifying formal systems.
A unique aspect of these calculi is that they let us define types which
are indexed by terms.
This can be a more intuitive method of encoding relationships
between terms and types within a specification than using predicates,
such as in Prolog.
To take a computational view, we interpret types as formulas, and proving
such formulas then reduces to to checking that a certain type
is inhabited.
The particular dependently typed $\lambda$-calculus we shall use in
this paper is called the Edinburgh Logical Framework or LF.
We describe this calculus below, then exhibit its use in specifying
relations and finally explain how it can be given an executable
interpretation.

\subsection{The Edinburgh Logical Framework}
There are three categories of LF expressions: kinds, type families
which are classified by kinds, and objects which are classified by
types.
Below, $x$ denotes an object variable, $c$ an object constant, and $a$
a type constant.
Letting $K$ range over kinds, $A$ and $B$ over types, and $M$ and $N$
over objects, the syntax of these terms are as follows:
\begin{tabbing}
  \=\qquad\=\quad\=\quad\ \=\kill
  \>\>$K$ \> $:=$ \> $\type\ |\ \typedpi{x}{A}{K}$ \\
  \>\>$A$ \> $:=$ \> $a\ |\ \typedpi{x}{A}{B}\ |\ \app{A}{M} $ \\
  \>\>$M$ \> $:=$ \> $c\ |\ x\ |\ \typedlambda{x}{A}{M}\ |\ \app{M}{N} $
\end{tabbing}
Both $\Pi$ and $\lambda$ are binders which assign a type to a variable
over the term.
The shorthand $A\ra P$ is used for $\typedpi{x}{A}{P}$ when $x$ does
not appear free in $P$.
Terms differing only in bound variable names are identified.
We use $U$ and $V$ below to stand ambiguously for types and
object expressions.
We write $\subst{U}{M_1/x_1,\ldots,M_n/x_n}$ to denote the capture avoiding
substitution of $M_1,\ldots,M_n$ for free occurrences of $x_1,...,x_n$
in $U$.

LF type family and object expressions are formed starting from a
signature $\Sigma$ that identifies constants together with their
kinds or types.
In addition, in determining whether or not an expression is
well-formed, we will need to consider contexts, denoted by $\Gamma$,
that assign types to variables.
The syntax for signatures and contexts is as follows:
\begin{tabbing}
  \=\qquad\=\quad\=\quad\ \=\kill
  \>\>$\Sigma$ \> $:=$ \> $\emptyctx \ |\ \Sigma,\oftype{a}{K}\ |\ \Sigma,\oftype{c}{A}$\\
  \>\>$\Gamma$ \> $:=$ \> $\emptyctx \ |\ \Gamma,\oftype{x}{A}$
\end{tabbing}
In what follows, the signature (which is user-defined)
will not change over the course of time.
Given this, for simplicity, we will leave this signature implicit in
our discussions.

Not all the LF expressions identified by the syntax rules above are
considered to be well-formed.
The following five forms of judgments are relevant to deciding the
ones that are:
\begin{gather*}
\Sigma~\sig \qquad \lfprove{}{\Gamma~\ctx} \\
\lfprove{\Gamma}{K~\kind} \qquad \lfprove{\Gamma}{\oftype{A}{K}} \qquad \lfprove{\Gamma}{\oftype{M}{A}}
\end{gather*}
The judgments on the first line assert, respectively, that $\Sigma$ is
a valid signature and that $\Gamma$ is a valid context, implicitly in
$\Sigma$.
The judgments on the second line assert that $K$ is a valid kind in
the context $\Gamma$, $A$ is a valid type of (valid) kind $K$ in
$\Gamma$, and $M$ is a valid object of (valid) type $A$ in $\Gamma$;
all these judgments also verify that the context $\Gamma$ and the
implicit signature $\Sigma$ are both valid.
In stating the rules for deriving these judgments, we shall make use
of an equality notion for expressions that is based on
$\beta$-conversion, \ie, the reflexive and transitive closure of a
relation that equates two expressions that differ only in that a
subexpression of the form $(\app{(\typedlambda{x}{A}{M})}{N})$  in one is
replaced by $\subst{M}{N/x}$ in the other.
We shall write $U^\beta$ for the $\beta$-normal form of an expression,
\ie, for an expression that is equal to $U$ and that does
not contain any subexpressions of the form
$(\app{(\typedlambda{x}{A}{M})}{N})$.
The rules for deriving the five different LF judgments are presented in
Figure~\ref{fig:lf-rules}.
Notice that we allow for the derivation of judgments of the form
$\lfprove{\Gamma}{\oftype{A}{K}}$ and
$\lfprove{\Gamma}{\oftype{M}{B}}$ only when $K$ and $B$ are in
$\beta$-normal form.
We also observe that such forms are not guaranteed to exist for all LF
expressions.
However, they do exist for well-formed LF expressions
\cite{harper93jacm}, a property that is ensured to hold for each
relevant LF expression by the premises of every rule whose conclusion
requires the $\beta$-normal form of that expression.

The notion of equality that we use for LF terms also includes
$\eta$-conversion, \ie, the congruence generated by the relation that
equates $\typedlambda{x}{A}{(\app{M}{x})}$ and $M$ if $x$ does not appear
free in $M$.
Observe that $\beta$-normal forms for the different categories of
expressions have the following structure
\begin{tabbing}
  \=\qquad\=\qquad\quad\ \=\kill  
  \>\>$Kind$ \> $\typedpi{x_1}{A_1}{\ldots\typedpi{x_n}{A_n}{Type}}$\\
  \>\>$Type$ \> $\typedpi{y_1}{B_1}{\ldots\typedpi{y_n}{B_m}{a\ M_1\ \ldots\ M_n}}$\\
  \>\>$Object$ \> $\typedlambda{x_1}{A_1}{\ldots\typedlambda{x_n}{A_n}{u\ M_1\ \ldots\ M_n}}$
\end{tabbing}
where $u$ is an object constant or variable and where the subterms and
subtypes appearing in the expression recursively have the same form.
We refer to the the part denoted by $a\ M_1\ \ldots\ M_n$ in a
type expression in such a form as its \emph{target} type and to
$B_1,\ldots,B_m$ as its \emph{argument} types.
Let $w$ be a variable or constant which appears in the well-formed
term $U$ and let the number of $\Pi$s that appear in the prefix of its
type or kind be $n$.
We say $w$ is {\it fully applied} if every occurrence of $w$ in $U$
has the form $\app{w}{M_1\ldots M_n}$.
A type of the form $\app{a}{M_1\ldots M_n}$ where $a$ is fully applied
is a {\it base type}.
We also say that $U$ is {\it canonical} if it is in normal form and every
occurrence of a variable or constant in it is fully applied.
It is a known fact that every well-formed LF expression is equal to
one in canonical form by virtue of
$\beta\eta$-conversion~\cite{harper93jacm}.

\subsection{Specifying Relations in LF}
LF can be used to formalize different kinds of rule based systems by
describing a signature corresponding to the system, as we now illustrate.
In presenting particular signatures, we will use a more
machine-oriented syntax for LF expressions: we write $\{x:U\}V$
for $\typedpi{x}{U}{V}$ and $[x:A]M$ for $\typedlambda{x}{A}{M}$.

The first example we consider is that of the natural number system.
To formalize this system we must, first of all, provide a
representation for the numbers.
This is easy to do: we pick a type corresponding to these numbers and
then provide an encoding for zero and the successor constructor.
The first three items in the signature shown in
Figure~\ref{fig:rel-ex} suffice for this purpose.
The next thing to do is to specify operations on natural numbers.
In LF we think of doing this through relations: thus, addition would
be specified as a relation between three numbers.
To describe relations we use \emph{dependent} types.
For example, the addition relation might be encoded as a
\emph{type} constant that takes three natural number \emph{objects} as
arguments.
The real interest is in determining when such a relation holds.
In rule based specifications this is typically done through inference
rules.
Thus, using the LF notation that we have just described, addition
might be defined by the rules
\begin{center}
\begin{tabular}{cc}
  \AxiomC{}
  \UnaryInfC{$plus~z~X~X$}
  \DisplayProof
  & \qquad\qquad
  \AxiomC{$plus~N~M~L$}
  \UnaryInfC{$plus~(s~N)~M~(s~L)$}
  \DisplayProof
\end{tabular}
\end{center}
in which tokens represented by uppercase letters constitute schema
variables.
In an LF specification, such rules correspond to object constants
whose target type is the representation of the rule's conclusion
and whose argument types are the types of the schema variables and the
representations of the premises.
As a concrete example, the object constants $plusZ$ and $plusS$
defined in Figure~\ref{fig:rel-ex} represent the two addition rules
shown.
The question of whether a relation denoted by a type holds now
becomes that of whether we can use the constants representing the
rules to construct an object expression of that type.
Thus, types function as formulas in an LF-style specification and
the provability of a formula becomes the question of type inhabitation.

We illustrate these ideas once more by using the example of lists of
natural numbers.
To represent such lists, we use the type $list$ and the object
constants $nil$ and $cons$ defined in Figure~\ref{fig:rel-ex}.
Now consider the append relation on these lists.
This relation is represented by the type constant $append$ that takes
three object-level expressions of type $list$ as arguments.
The rules for proving this relation are the following
\begin{center}
\begin{tabular}{cc}
  \AxiomC{}
  \UnaryInfC{$append~nil~L~L$}
  \DisplayProof
  &
  \AxiomC{$append~L~M~N$}
  \UnaryInfC{$append~(cons~X~L)~M~(cons~X~M)$}
  \DisplayProof
\end{tabular}
\end{center}
Following the structure described earlier, the object constants
$appNil$ and $appCons$ shown in Figure~\ref{fig:rel-ex} represent
these rules.
\begin{figure}
  \begin{tabbing}
    \=\qquad\quad\=\qquad\=\kill
    \>$nat$\>$: type.$\\
    \>$z$\>$: nat.$\\
    \>$s$\>$: nat\ra nat.$\\
    \\
    \>$plus$\>$: nat\ra nat\ra nat\ra type.$\\
    \>$plusZ$\>$: \{x:nat\}plus~z~x~x.$\\
    \>$plusS$\>$: \{l:nat\}\{m:nat\}\{n:nat\}plus~l~m~n\ra$\\
    \>\>\>$plus~(s~l)~m~(s~n).$
  \end{tabbing}
  \begin{tabbing}
  \=\qquad\qquad\=\ \ \ \=\qquad\=\kill
  \>$list$\>$:$\>$type.$\\
  \>$nil$\>$:$\>$list.$\\
  \>$cons$\>$:$\>$nat\ra list\ra list.$\\
  \\
  \>$append$\>$:$\>$list\ra list\ra list\ra type.$\\
  \>$appNil$\>$:$\>$\{l:list\}append~nil~l~l.$\\
  \>$appCons$\>$:$\> $\{x:nat\}\{l:list\}\{m:list\}\{n:list\}$ \\
  \>\>\>\> $append~l~m~n\ra$\\
  \>\>\>\> $append~(cons~x~l)~m~(cons~x~n).$
  \end{tabbing}
\nocaptionrule  \caption{An example of specifications in LF}
  \label{fig:rel-ex}
\end{figure}

\subsection{Logic Programming}

The Twelf system gives LF specifications a logic programming
interpretation.
Computation is initiated in Twelf by presenting it with a type.
Such a type, as we have explained earlier, corresponds to a formula
and the task is to find a proof for it or, more precisely, to find an
inhabitant for the provided type.

The search problem is actually better viewed as that
of checking if a given object expression $M$ has a given type $A$;
this formulation subsumes the case where only the type is given
because we allow $M$ to contain variables that may become instantiated
as the search progresses.
In the simple case $A$ is a base type.
Here, computation proceeds by looking for an object declaration
\[ c : \{x_1:B_1\}\ldots\{x_n:B_n\}A' \]
in the signature at hand and checking if there are object expressions
$M_1,\ldots,M_n$ such that $\subst{A'}{M_1/x_1,\ldots,M_n/x_n}$ is
equal to $A$.
If this is the case and if it is also the case that $M$ and
$c\ M_1\ \ldots\ M_n$ can be unified, then the task reduces, recursively, to
checking if $M_i$ has the type $B_i$ for $1 \leq i \leq n$.
In this model of computation, the types associated with object
constants in a signature are often referred to as \emph{clauses} and
the process of picking an object declaration and trying to use it to
solve the inhabitation question is referred to as \emph{backchaining}
on a clause.

In the more general case, $A$ may not be a base type, \ie~it may
actually have the structure $\{x_1:A_1\}\ldots\{x_m:A_m\}B$ where $B$
is a base type.
In this case, we first transform the task to trying to show that the
object expression $M\ x_1\ \ldots\ x_m$ has type $B$ where we
treat $x_1,\ldots,x_m$ as new constants of type $A_1,\ldots,A_m$,
respectively, that are dynamically added to the signature.

For a concrete example of this behavior, let our
signature be the specification of append from
Figure~\ref{fig:rel-ex} and let our goal be to construct a term $M$
such that
$$\lfprove{}{\oftype{M}{append~(cons~z~nil)~nil~(cons~z~nil)}}$$
is derivable.
We can match this type with the target type of $appCons$ and we are then
left with finding a term $N$ such that
$$\lfprove{}{\oftype{N}{append~nil~nil~nil}}$$
is derivable.
Notice that this step also results in $M$ being
instantiated to $appCons\ z\ nil\ nil\ nil\ N$.
The type in the new goal of course matches that of $appNil$, resulting
in $N$ being instantiated to $appNil~nil$ and $M$ correspondingly
being instantiated with the expression
$$appCons~z~nil~nil~nil~(appNil~nil)$$
We have, at this point determined that this object expression inhabits
the type
$append~(cons~z~nil)~nil~(cons~z~nil)$.

  \section{Specifications in Predicate Logic}
\label{sec:hohh}
%
\begin{figure*}
\begin{center}
\begin{tabular}{ccc}
    \AxiomC{}
    \RightLabel{\topGoal}
    \UnaryInfC{$\sigsequent{\Xi}{\Gamma}{\top}$}
    \DisplayProof

&

    \AxiomC{$\sigsequent{\Xi}{\Gamma \cup\{D\}}{G}$}
    \RightLabel{\impGoal}
    \UnaryInfC{$\sigsequent{\Xi}{\Gamma}{D \supset G}$}
    \DisplayProof

&

    \AxiomC{$c \notin \Xi \quad \sigsequent{\Xi \cup
        \{c\}}{\Gamma}{G[c/x]}$}
    \RightLabel{\allGoal}
    \UnaryInfC{$\sigsequent{\Xi}{\Gamma}{\forallx{x}{G}}$}
    \DisplayProof

\end{tabular}

\medskip

\begin{tabular}{cc}
    \AxiomC{$D\in \Gamma \quad \focusedsigseq{\Xi}{\Gamma}{A}{D}$}
    \RightLabel{\atomicGoal}
    \UnaryInfC{$\sigsequent{\Xi}{\Gamma}{A}$}
    \DisplayProof

&

    \AxiomC{}
    \RightLabel{\initial}
    \UnaryInfC{$\focusedsigseq{\Xi}{\Gamma}{A}{A}$}
    \DisplayProof
\end{tabular}

\medskip
\begin{tabular}{cc}
    \AxiomC{$t\ \mbox{\rm is a}\ \Xi\mbox{\rm -term} \quad
      \focusedsigseq{\Xi}{\Gamma}{A}{\subst{D}{t/x}}$}
    \RightLabel{\allClause}
    \UnaryInfC{$\focusedsigseq{\Xi}{\Gamma}{A}{\forallx{x}{D}}$}
    \DisplayProof

&
    \AxiomC{$\sigsequent{\Xi}{\Gamma}{G} \quad
      \focusedsigseq{\Xi}{\Gamma}{A}{D}$}
    \RightLabel{\impClause}
    \UnaryInfC{$\focusedsigseq{\Xi}{\Gamma}{A}{G \supset D}$}
    \DisplayProof
\end{tabular}
\end{center}
\nocaptionrule \caption{Derivation rules for the \hhf\ logic}
  \label{fig:hhfrules}
\end{figure*}

Another approach to specification uses a predicate logic,
where relations are encoded as predicates rather than in types.
The idea of executing the specifications then corresponds to
constructing a proof for chosen formulas in the relevant logic.
To yield a sensible notion of computation, the specifications must
also be able to convey information about how a search for a proof
should be conducted.
Not all logics are suitable from this perspective.
Here we describe the logic of higher-order hereditary Harrop formulas
that does have an associated computational interpretation and that, in
fact, is the basis for the programming language $\lambda$Prolog
\cite{nadathur88iclp}.
We present the syntax of the formulas in this logic in the first
subsection below and then explain their computational interpretation.
The \hhf\ logic will be the target for the translation of Twelf that
is the focus of the rest of the paper.

\subsection{Higher-order hereditary Harrop formulas}
The \hhf~logic is based on Church's Simple Theory of Types 
\cite{church40}.
The expressions of this logic are those of a simply typed
$\lambda$-calculus.
The types are constructed from the atomic type $o$ of
propositions and a finite set of other atomic types by using the
function type constructor $\ra$.
There are assumed to be two sets of atomic expressions, one
corresponding to variables and the other to constants, in which each
member is assumed to have been given a type.
All typed terms can be constructed from these typed sets of
constants and variables by application and $\lambda$-abstraction.
As in LF, terms differing only in bound variable names are identified.
The notion of equality between terms is further enriched by $\beta$-
and $\eta$-conversion.
When we orient these rules and think of them as reductions, we are
assured in the simply typed setting of the existence of a
unique normal form for every well-formed term under these reductions.
Thus, equality between two terms becomes the same as the identity of
their normal forms.
For simplicity, in the remainder of this paper we will assume that all
terms have been converted to normal form.
We use $\subst{t}{s_1/x_1,\ldots,s_n/x_n}$ to denote the capture
avoiding substitution of the terms $s_1,\ldots,s_n$ for free
occurrences of $x_1,...,x_n$ in $t$.

Further qualifications are required to introduce logic into this
setting.
First, the constants mentioned above are divided into the categories
of logical and non-logical constants.
Next, we restrict the constants so that only the logical constants can
have argument types containing the type $o$.
Finally, we limit the logical constants to the following:
\begin{tabbing}
  \=\qquad\=\quad\ \=\qquad\quad\ \=\kill
  \>\>$\top$ \> of type \> $o$\\
  \>\>$\supset$ \> of type \> $o\ra o\ra o$\\
  \>\>$\Pi$ \> of type \> $(\tau\ra o)\ra o$ for each valid type $\tau$
\end{tabbing}
$\Pi$ denotes universal quantification, and the shorthand
$\forallx{x}{F}$ is used for $\Pi(\lambdax{x}{F})$.

The set of non-logical constants is typically called the signature,
and as mentioned $o$ cannot appear in the type of any argument of
these constants.
However, $o$ is allowed as the target type for nonlogical constants.
Constants with target type $o$ are called predicates; those with any
other target type are called constructors.

For a nonlogical constant $c$ of type $\tau_1\ra\ldots\ra\tau_n\ra o$
and terms $t_1,\ldots,t_n$ of type $\tau_1,\ldots,\tau_n$, we call the
term $c~t_1\ldots\ t_n$ of type $o$ an \ital{atomic formula}.
Using the set of logical constants, we construct sets of $G$ and
$D$-formulas from the set of atomic formulas.
The syntax of these two sets is the following:
\begin{tabbing}
   \=\qquad\=\quad\=\quad\ \=\kill
   \>\>$G$ \> $:=$ \> $\top\ |\ A\ |\ D \supset G\ |\ \forallx{x}{G}$\\
   \>\>$D$ \> $:=$ \> $A\ |\ G \supset D\ |\ \forallx{x}{D}$
\end{tabbing}
where $A$ denotes an atomic formula.

The $D$ formulas described above are also called higher-order
hereditary formulas.
A specification in this setting consists of a set of such formulas.
To illustrate how such specifications may be constructed in practice,
let us consider the encoding of the append relation on lists of
natural numbers.
The first step in formalizing this relation is to describe a
representation for the data objects in its domain.
Towards this end, we introduce two atomic types, $nat$ and $list$.
Our signature should then identify the obvious constructors with each
of these types:
\begin{tabbing}
\=\qquad\=\qquad\ \=\qquad\quad\ \=\kill
\>\>$z$ \> of type \> $nat$\\
\>\>$s$ \> of type \> $nat\ra nat$\\
\>\>$nil$  \> of type \> $list$\\
\>\>$cons$ \> of type \> $nat\ra list\ra list$
\end{tabbing}
As a concrete example, the list that has $0$ and $1$ as its elements
would be represented by the term
$(\app{\app{cons}{z}}{(\app{\app{cons}{(\app{s}{z})}}{nil})})$.

The append relation will now be encoded via a {\it predicate}
constant, \ie, a non-logical constant that has $o$ as its target type.
In particular, we might use the constant $append$ that has the type
\[list\ra list\ra list\ra o\]
for this purpose.
To define the relation itself, we might use the following two
$D$-formulas:
\begin{center}
  \begin{tabular}{l}
  $\forallx{l}{(append~nil~l~l)}$\\
  $\forallx{x}{\forallx{l_1}{\forallx{l_2}{\forallx{l_3}{(append~l_1~l_2~l_3)\supset}}}}$\\
  $\qquad\qquad\qquad\qquad (append~(cons~x~l_1)~l_2~(cons~x~l_3))$
  \end{tabular}
\end{center}
These formulas, that are also often referred to as the {\it clauses}
of a specification or program, can be visualized as defining the
append relation by recursion on the structure of the list that is its
first argument.
The first formulas treats the base case, when this list is empty.
The second formula treats the recursive case; the conclusion of the
implication is conditioned by the relation holding in the
case where the first argument is a list of smaller size.
This pattern, of using universal quantifications over atomic formulas
to treat the base cases of a relation and such quantifications over
formulas that have an implication structure to treat the recursive
cases is characteristic of relational specifications in the
\hhf\ logic.

\subsection{Logic Programming}
The computational interpretation of the \hhf\ logic consists of
thinking of a  collection of $D$-formulas as a program against which
we can solve a $G$-formula.
More formally, computation in this setting amounts to attempting to
construct a derivation for a sequent of the form
$\sequent{\Xi;\mathcal{P}}{G}$,
where $\Xi$ is a signature, $\mathcal{P}$ is a set
of program clauses, and $G$ a goal formula.
The computation that results from such a sequent consists of first
decomposing the goal $G$ in a manner determined by the logical
constants that appear in it and then, once $G$ has been broken up into
its atomic components, picking a formula from $\mathcal{P}$ and using
this to solve the resulting goals.

The precise derivation rules for the \hhf~logic are given in
Figure~\ref{fig:hhfrules}.
These rules can be understood as follows.
In a sequent of the form $\sequent{\Xi;\mathcal{P}}{G}$, if $G$ is
not an atomic formula, then it must have one of the forms $\top$,
$D\supset G'$ or $\forall x.G'$.
The first kind of goal has an immediate solution.
In the second case, we extend the logic program $\mathcal{P}$ with ${D}$
and continue search with $G'$ as the new goal formula.
In the last case, \ie, when $G$ is of the form $\forall x.G'$, we
expand $\Xi$ with a new constant $c$ and the new goal becomes
$G'[c/x]$.
Once we have arrived at an atomic formula $A$, we pick a clause from
$\mathcal{P}$ whose head eventually ``matches'' $A$, spawning off new
goals to solve in the process.
The exact manner in which this kind of simplification of atomic goals
takes place is determined by the last four rules in
Figure~\ref{fig:hhfrules}.

A special case for treating atomic goals arises when the clause
selected from the program has the structure
\[\forall x_1.(F'_1\supset\ldots\supset\forall x_n.(F'_n\supset
A')\ldots),\] and where it is the case that for terms $t_1,\ldots,t_n$
of correct type, $A=A'[t_1/x_1,\ldots,t_n/x_n]$.
The effect of the sequence of rule applications that results in this
case is reflected in the following derived rule
  \begin{center}
    \AxiomC{$\sigsequent{\Xi}{\Gamma}{F_1}\qquad \ldots \qquad
      \sigsequent{\Xi}{\Gamma}{F_n}$}
    \RightLabel{\bcGoal}
    \UnaryInfC{$\sigsequent{\Xi}{\Gamma}{A}$}
    \DisplayProof
  \end{center}
in which $F_i=F'_i[t_1/x_1,\ldots,t_i/x_i]$ for $0<i\leq n$. We
shall find this rule, which we have labeled $\bcGoal$ for obvious
reasons, useful in the analyses that appears in later sections.

\begin{figure}
\begin{tabbing}
\=\qquad\qquad\=\kill
  \>$kind~nat~type.$\\
  \>$kind~list~type.$\\
\\
  \>$type~z~nat$\\
  \>$type~s~(nat\ra nat).$\\
  \>$type~nil~list.$\\
  \>$type~cons~(nat\ra list\ra list).$\\
\\
  \>$pi~(L\backslash~(append~nil~L~L)).$\\
  \>$pi~x\backslash~(pi~L1\backslash~(pi~L2\backslash~(pi~L3\backslash~$\\
  \>\>$(append~L1~L2~L3~=>$\\
  \>\>$append~(cons~X~L1)~L2~(cons~X~L3))))).$
\end{tabbing}
\nocaptionrule  \caption{An example of a $\lambda$Prolog program}
  \label{fig:app-sig}
\end{figure}

The $\lambda$Prolog language can be viewed as a programming rendition
of the \hhf~logic that we have discussed here.
In $\lambda$Prolog, the user can introduce new atomic types through
declarations that begin with the keyword $kind$ and new constructors
by using declarations that begin with the keyword $type$.
Examples of such declarations appear in Figure~\ref{fig:app-sig}.
A complete program consists not only of such declarations that
identify the signature, but also of $D$-formulas that define
relations.
In the concrete syntax of $\lambda$Prolog, abstraction is written as
the infix symbol \verb+\+, \ie, the expression $\lambda x\, F$ is
  rendered as \verb+x\ F+.
Moreover, the logical constants
$\top$, $\Pi$ and $\supset$ are written
as {\tt true}, {\tt pi} and {\tt =>} respectively.
Another option for expressing {\tt G => D} is the notation {\tt D :-
  G}.
Several of these aspects of $\lambda$Prolog syntax are illustrated in
Figure~\ref{fig:app-sig} through the presentation of clauses defining
the {\it append} relation.

The $\lambda$Prolog language has been given an efficient
compilation-based implementation in the Teyjus system.
One of the goals of our work is to leverage this implementation in
providing also an efficient treatment of Twelf programs.

  \section{A Naive Translation}
\label{sec:naive}

\begin{figure}
  \centering
   \[ \begin{array}{c}
   \phi(A) := \lfobject \ \text{when $A$ is a base type} \\
   \phi(\typedpi{x}{A}{P}) := \phi(A)\ra \phi(P)
   \quad\quad
   \phi(\type) := \lftype
   \end{array} \]
\[\begin{array}{c}
    \encTerm{u\ M_1 \ldots M_n} := u \ \encTerm{M_1}\ldots\encTerm{M_n} \\
\quad\quad
    \encTerm{x\ M_1 \ldots M_n} := x \ \encTerm{M_1}\ldots \encTerm{M_n}\\
\quad\quad
    \encTerm{\typedlambda{x}{A}{M}} := \lambda^{\phi(A)} x. \encTerm{M}
\end{array} \]
  \begin{align*}
    \enc{\typedpi{x}{A}{B}} :=&\
      \lambda M.~
      \forall x.~ (\app{\enc{A}}{x}) \supset (\enc{B}~(\app{M}{x})) \\
    \enc{A} :=&\
      \lambda M.~
      \hastype{M}{\encTerm{A}}
      \ \text{where $A$ is a base type}
  \end{align*}
\nocaptionrule  \caption{Encoding of types, objects, and
             translation of LF judgments to \hhf}
  \label{fig:naive-translation}
\end{figure}

We present in this section a simple translation of LF specifications
into \hhf~specifications.
This translation is taken from \cite{snow10ppdp} that builds on earlier
work due to Felty~\cite{felty90cade}.
After presenting the translation, we will prove a correspondence
between its source and target.
This property will ensure that reasoning based on the translation will
correctly follow reasoning based on the original specification.
In this way, we know that constructing a \hhf~proof of some judgment
is equivalent to finding a derivation in LF.
Unfortunately, the simple translation produces \hhf~formulas that
contain a lot of redundant information related to type checking that
can result in quite inefficient proof search behavior.
We highlight this issue towards developing a better translation in the
next section.

\subsection{The Translation}
We have previously seen two methods for specifying append, in
Section~\ref{sec:lf} a dependently-typed calculus was used and in
Section~\ref{sec:hohh} we utilized a relational style.
Similarities between these two styles should have become apparent from
this simple example.
The signature we defined consisted of expressions which are
essentially the same between LF and the simply typed $\lambda$ calculus.
Differences appear when defining dependencies between objects and
types.
In LF these relations are defined in the types and so we defined
objects $appNil$ and $appCons$.
\hhf~is simply typed, and so relations are encoded using predicates and
$D$-formulas are constructed to define exactly when the relation
holds.
There is then, a clear connection between the dependent types in LF,
and the program clauses in \hhf.
The closeness of these two approaches is important in determining a
translation from LF to \hhf~specifications.

As we have seen in Section~\ref{sec:lf}, the goal of proof search in
Twelf is to determine if an object of a particular type can be
constructed.
We will mimic this situation in $\lambda$Prolog by
examining if we can construct a proof for an \hhf~formula
that is obtained from the LF type.
The translation presented by Felty relies on having in hand both the
LF type and the LF object, but this is obviously too much to expect if
proof search is intended to be the main focus.
To overcome this difficulty, Snow~\etal adapted Felty's translation so
that it was based solely on the type \cite{snow10ppdp}; the LF object
is then uncovered incrementally by proof search in the \hhf~logic from
the corresponding specification.

This translation, which is presented in
Figure~\ref{fig:naive-translation}, uses a two step process.
In the first step a coarse mapping is described that takes both LF types
and objects to \hhf~terms.
More specifically, \hhf~terms of type $\lftype$ and $\lfobject$ are
used to represent LF objects of base kinds and types respectively.
The mapping $\phi$ then identifies an \hhf~type with
each arbitrary LF kind and object.
Finally, an LF object $M$ of type $A$ is encoded by the \hhf~term
$\encTerm{M}$ of type $\phi(A)$, and respectively, the type $B$ of
kind $K$ is encoded by the \hhf~term $\encTerm{B}$ of type $phi(K)$.
This simple mapping clearly loses much of the dependency
information available in the original LF types and kinds.
In the second pass, we recover the lost information by making use of
an \hhf~predicate $hastype$ of type $\lfobject\ra\lftype\ra o$:
$\hastype{X}{T}$ is to hold exactly when $X$ is the encoding of some
LF term $M$ of a base LF type whose encoding is $T$.
In more detail, using this predicate, we translate each LF type $A$
into an \hhf~predicate term $\enc{A}$ that is intended to take the
encodings of LF objects as arguments.
Interpreting $\enc{\oftype{M}{A}}$ as $(\app{\enc{A}}{\encTerm{M}})$
and using this to describe also the translations $\enc{\Gamma}$ and
$\enc{\Sigma}$ of LF contexts and signatures, we expect our
translations to be such that, for a suitable \hhf~signature $\Xi$,
\[\iprove{\Xi; \enc{\Sigma}, \enc{\Gamma}}{(\app{\enc{A}}{\encTerm{M}})}\]
is derivable in the
\hhf~logic just in the case that $\lfprove{\Gamma}{\oftype{M}{A}}$ is
a valid LF judgment.\footnote{To translate LF signatures in their
  entirety, we also have to describe a translation of kinds. However,
  these translations will not be used in the derivations in \hhf\ and
  so we make them explicit.}

\begin{figure}
  \begin{tabbing}
  \=\qquad\=\qquad\qquad\=\kill
  \> $nat : \lftype$\\
  \> $z : \lfobject$\\
  \> $s : \lfobject \ra \lfobject$\\
  \> $list : \lftype$\\
  \> $nil : \lfobject$\\
  \> $cons : \lfobject \ra \lfobject \ra \lfobject$\\
  \> $append : \lfobject \ra \lfobject \ra \lfobject \ra \lftype$\\
  \> $appNil : \lfobject \ra \lfobject$\\
  \> $appCons : \lfobject \ra \lfobject \ra \lfobject \ra \lfobject \ra \lfobject \ra \lfobject$\\
  \\
  \> $\hastype{z}{nat}$ \\
  \> $\forall n.~ \hastype{n}{nat} \supset \hastype{(s~n)}{nat}$\\
  \> $\hastype{nil}{list}$ \\
  \> $\forall n.~ \hastype{n}{nat} \supset \forall l.~ \hastype{l}{list} \supset$\\
  \>\> $\hastype{(cons~n~l)}{list}$ \\
  \\
  \> $\forall l.~ \hastype{l}{list}\supset \hastype{(appNil~l)}{(append~nil~l~l)}$ \\
  \> $\forall x.~ \hastype{x}{nat} \supset \forall l.~ \hastype{l}{list} \supset$\\
  \>\> $\forall k.~ \hastype{k}{list} \supset \forall m.~ \hastype{m}{list} \supset$\\
  \>\> $\forall a.~ \hastype{a}{(append~l~k~m)} \supset $\\
  \>\> $hastype~(appCons~x~l~k~m~a)$\\
  \>\>\> $(append~(cons~x~l)~k~(cons~x~m))$
  \end{tabbing}
\nocaptionrule  \caption{Translation of the LF specification for $append$}
  \label{fig:append-translation}
\end{figure}

Figure~\ref{fig:append-translation} illustrates the translation of an
LF signature into an \hhf~program using the
example LF signature of append shown in Figure~\ref{fig:rel-ex}.
We would like to use the \hhf\ ($\lambda$Prolog) program that results
from such a translation as the basis for responding to inhabitation
questions raised relative to Twelf specifications.
The ambient \hhf~program and signature in the \hhf~sequents that we
have to consider in this setting arise from LF signatures that we are
already leaving implicit.
We will therefore also elide these parts of the \hhf~sequent, writing
$\iprove{\Xi; \enc{\Sigma},
  \enc{\Gamma}}{(\app{\enc{A}}{\encTerm{M}})}$
more simply as
$\iprove{\enc{\Gamma}}{(\app{\enc{A}}{\encTerm{M}})}$,
mentioning explicitly at most those parts of the \hhf~signature that
result from the use of the \allGoal~rule during proof search.

The following theorem makes precise our informal description of the
property of our translation and also provides the basis for using
\hhf~proof search in answering LF queries.

\begin{theorem'}\label{thm:naive}
  Let $\Gamma$ be a well-formed canonical LF context and let $A$ be a
  canonical LF type such that $\lfprove{\Gamma}{\oftype{A}{\type}}$
  has a derivation.
  If $\lfprove{\Gamma}{\oftype{M}{A}}$ has a derivation for a
  canonical object $M$, then there is a derivation of
  $\iprove{\enc{\Gamma}}{\enc{\oftype{M}{A}}}$.
  Conversely, if $\iprove{\enc{\Gamma}}{(\app{\enc{A}}{M'})}$ for an
  arbitrary \hhf~term $M'$, then there is a canonical LF object $M$
  such that $M'=\encTerm{M}$ and $\lfprove{\Gamma}{\oftype{M}{A}}$ has
  a derivation.
\end{theorem'}
The proof of this theorem can be found in~\cite{snow10ppdp}.
To summarize the proof, the completeness argument proceeds by
induction on the derivation of
$\lfprove{\Gamma}{\oftype{M}{A}}$ to show how to construct a
derivation for $\iprove{\enc{\Gamma}}{\enc{\oftype{M}{A}}}$.
Similarly, for soundness it uses induction on the derivation of
$\iprove{\enc{\Gamma}}{(\app{\enc{A}}{M'})}$ to extract from $M'$ an LF
object $M$ of the required type.

\subsection{Some Issues With the Translation}
The translation described here has been shown correct.
However, because LF expressions contain a lot of redundant
information, and because of the context in which we want to use the
translation, it is possible to produce a version that is more
optimized for proof search.
A key fact to bear in mind is that when we consider judgments of the form
$\lfprove{\Gamma}{\oftype{M}{A}}$ in the setting of logic programming,
we would have already verified that $A$ is a valid type.
This knowledge gives us additional typing related information.
For example, suppose that
\[A=append~nil~(cons~z~nil)~(cons~z~nil).\]
If we know that $A$ is a valid type, then clearly $(cons~z~nil)$ must
be of type $list$.
In fact, looking at the \appfam~rule tells us that a derivation of
$\lfprove{\Gamma}{\oftype{A}{\type}}$ must contain a derivation of
$\lfprove{\Gamma}{\oftype{(cons~z~nil)}{list}}$.
Thus, in deriving the \hhf~goal
\[
\hastype{M}{(append~nil~(cons~z~nil)~(cons~z~nil))}
\]
it is unnecessary to show that $(\hastype{(cons~z~nil)}{list})$ holds
as the translation of the type of $appCons$ that is shown in
Figure~\ref{fig:append-translation} requires us to do.

Removing tests like those above that arise from binders in LF types
would certainly simplify the \hhf~specification and would thereby
allow for more efficient proof search.
However, not all such binders can be ignored in the translation: some
of them also play a role in addressing inhabitation questions and are
not just relevant to type checking.
For example, consider the (well-formed) type
\[append~(cons~z~nil)~(cons~z~nil)~(cons~z~nil).\]
To form an object of this type based on the $appCons$ constructor, we
need to have in hand an object of type
\[append~z~(cons~z~nil)~nil.\]
Thus, the translation of the type of $appCons$ in whose binder this
type occurs must preserve the subgoal corresponding to finding such an
object.
Clearly then, we need some method of determining which tests are
redundant and so can be correctly removed and which must be
preserved.

  \section{Improving the Translation}
\label{sec:refined}
%
\begin{figure}
\centering
\renewcommand{\arraystretch}{3}
\begin{tabular}{c}
    \AxiomC{$\termUV{dom(\Gamma); \cdot; x}{A_i}$ for some $A_i$}
    \RightLabel{\ruvappt}
    \UnaryInfC{$\formulaUV{\Gamma; x}{c \vec{A}}$}
\DisplayProof
\\
  \AxiomC{$\formulaUV{\Gamma, \oftype{y}{A}; x}{B}$}
  \RightLabel{\ruvpit}
  \UnaryInfC{$\formulaUV{\Gamma; x}{\typedpi{y}{A}{B}}$}
\DisplayProof
\\
  \AxiomC{$\formulaUV{\Gamma_1;x}{B}$}
  \AxiomC{$\formulaUV{\Gamma_1,\oftype{y}{B},\Gamma_2;y}{A}$}
  \RightLabel{\ruvctxt}
  \BinaryInfC{$\formulaUV{\Gamma_1,\oftype{y}{B},\Gamma_2;x}{A}$}
\DisplayProof
\\
  \AxiomC{$y_i \in \delta$ for each $y_i \quad\ y_i$ distinct}
  \RightLabel{\ruvinito}
  \UnaryInfC{$\termUV{\Gamma; \delta; x}{\app{x}{\vec{y}}}$}
\DisplayProof
\\
  \AxiomC{$y \notin \Gamma$ and $\termUV{\Gamma; \delta; x}{M_i}$ for some $i$}
  \RightLabel{\ruvappo}
  \UnaryInfC{$\termUV{\Gamma; \delta; x}{\app{y}{\vec{M}}}$}
\DisplayProof
\\
  \AxiomC{$\termUV{\Gamma; \delta, y; x}{M}$}
  \RightLabel{\ruvabso}
  \UnaryInfC{$\termUV{\Gamma; \delta; x}{\typedlambda{y}{A}{M}}$}
\DisplayProof
\end{tabular}
\nocaptionrule  \caption{Strictly occurring variables in types and objects}
  \label{fig:ruvs}
\end{figure}

The redundancy issue highlighted in the previous section can
be rephrased as follows.
We are interested in translating an LF type of the form
$\typedpi{x_1}{A_1}{\ldots\typedpi{x_n}{A_n}{B}}$
into an \hhf~clause that can be used to determine if a type $B'$ can
be viewed as an instance $\subst{B}{M_1/x_1,\ldots,M_n/x_n}$ of the
target type $B$.
This task also requires us to show that $M_1,\ldots,M_n$ are
inhabitants of the types $A_1,\ldots,A_n$; in the naive translation,
this job is done by the $hastype$ formulas pertaining to $x_i$ and
$A_i$ that appear in the body of the \hhf~clause produced for the overall
type.
However, particular $x_i$ may occur in $B$ in a manner that already
makes it clear that the term $M_i$ that replace them in any instance of
$B$ must possess such a property.
What we want to do, then, is characterize such occurrences of $x_i$ so
that we can avoid having to include an inhabitation check in the \hhf~clause.

In this section, we define a strictness condition for variable
occurrences and, hence, for variables that possesses this kind of
property.
By using this condition, we can simplify the translation of a type
into an \hhf~clause without losing accuracy.
In addition to efficiency, such a translation also produces a result
that bears a much closer resemblance to the LF type from which it
originates.
The correctness of this new translation is shown using lemmas
about this strictness condition.

\subsection{The Strictness Property and Redundancies in Types}\label{sec:refined:lf}
\begin{figure}
  \begin{tabbing}
  \=\qquad\=\qquad\=\kill
  \>$b~:~(nat\ra nat)\ra type.$\\
  \>$c~:~\{w_1:nat\ra nat\}$\\
  \>\>$\{w_2:(nat\ra nat)\ra nat\ra nat\}$\\
  \>\>$b~(w_2~w1)\ra type.$\\
  \>$d~:~\{w_1:nat\ra nat\}$\\
  \>\>$\{w_2:(nat\ra nat)\ra nat\ra nat\}$\\
  \>\>$(\{z:b~(w2~w1)\}~c~w_1$\\
  \>\>\>$([w:nat\ra nat][y:nat]~(w_2~w_1)~(w~y))~z)\ra$\\
  \>\>\>$type.$\\
  \>$f~:~\{x:nat\ra nat\}$\\
  \>\>$\{y:\{z:b~x\}~c~([y:nat]~y)$\\
  \>\>\>$([w:nat\ra nat][y:nat]~x~(w~y))~z\}$\\
  \>\>$d~([y:nat]~y)~([w:nat\ra nat][y:nat]~x~(w~y))~y.$
  \end{tabbing}
\nocaptionrule  \caption{An example motivating the strictness condition}
  \label{fig:strict-ex}
\end{figure}
To understand the intuition underlying the strictness condition on
variable occurrences and its relevance to type checking, take as an
example the signature in Figure~\ref{fig:strict-ex}.
The main focus in this example is on the constant $f$ and its type;
the other declarations are included because they are needed in
constructing the type associated with $f$.
Substituting $t_1$ and $t_2$ for $x$ and $y$ respectively provides an instance 
of the target type of $f$ which has the form
\[d~([y:nat]~y)~([w:nat\ra nat][y:nat]~t_1~(w~y))~t_2.\]
Suppose we know that this is a valid type.
Then we would already know that $t_2$ has the type
\[\{z:b~t_1\}c~([y:nat]~y)~([w:nat\ra nat][y:nat]~t_1~(w~y))~z\]
and hence would not need to check this explicitly.
The fact that $t_2$ has this type follows from looking at its occurrence in
the type known to be valid and noting that the checking of the
type of $f$ has already established that any instance of the third
argument of $d$ in this setting must have as its type the
corresponding instance of the type of $y$:
\[\{z:b~x\}c~([y:nat]~y)~([w:nat\ra nat][y:nat]~x~(w~y))~z.\]
Analyzing this more closely, we see that the critical contributing
factors (to $t_2$ occurring in this way in the type) are that the
path down to the occurrence of $y$ is \emph{rigid}, \ie, it cannot be
modified by substitution and $y$ is not applied to arguments in a way
that could change the structure of the expression substituted for it.
These properties were formalized in a notion of strictness in \cite{snow10ppdp},
there inappropriately referred to as rigidity.

The criterion described in \cite{snow10ppdp} actually fails to
recognize some further cases in which dynamic type checking can be
avoided.
To understand this, consider the occurrence of $x$ in the target type
of $f$.
This occurrence appears applied to an argument that could end up ``hiding'' the
actual structure of any instantiation of $x$.
We see this concretely in the instance
\[d~([y:nat]~y)~([w:nat\ra nat][y:nat]~t_1~(w~y))~t_2.\]
considered earlier; we know something about the type of the term
resulting from $t_1~(w~y)$, but cannot conclude anything about the
type of $t_1$ itself from this.
Thus, this occurrence of $x$ is correctly excluded by the strictness
condition presented in \cite{snow10ppdp}.

Observe, however, that $x$ has another occurrence in the type of $f$,
in particular, in the type of the argument $y$.
Further, because this argument $y$ occurs strictly in the instantiated
target type, we would have statically checked its validity.
Looking at that type, which is
\[\{z:b~t_1\}c~([y:nat]~y)~([w:nat\ra nat][y:nat]~t_1~(w~y))~z,\]
we would know that $b~t_1$ is well formed and therefore $t_1$ is
an inhabitant of the expected type.

In summary, it seems possible to extend the strictness condition
recursively while preserving its utility in recognizing redundancy in
type checking. 
We consider occurrences of bound variables to be strict
in the overall type if they are strict in the types of other bound
variables that occur strictly in the target type.
The relation defined in Figure~\ref{fig:ruvs} formalizes this idea.
Specifically, we say that the bound variable $x_i$ occurs strictly in
the type $\typedpi{x_1}{A_1}{\ldots\typedpi{x_n}{A_n}{B}}$ if it is
  the case that
\[\formulaUV{\cdot;
  x}{\typedpi{x_1}{A_1}{\typedpi{x_{i-1}}{A_{i-1}}{\typedpi{x_{i+1}}{A_{i+1}}{\typedpi{x_n}{A_n}{B}}}}}\]
holds.

In the lemmas that follow, we formally prove the relationship between this
notion and redundancy in type checking that we have discussed above.

\begin{lemma'}\label{lem:strictterms}
Let $N_1,\ldots,N_n$ be LF objects and $\Gamma$,
$\Gamma_0$, and
$\Delta$ be LF contexts where $\Gamma_0=\oftype{x_1}{B_1},\ldots,\oftype{x_n}{B_n}$.
Further, let $M$ be an LF object, $A$ an LF type and $\delta\subseteq
dom(\Delta)$.
Finally, suppose for some $i$ there are derivations of
\begin{enumerate}
\item $\termUV{x_1,\ldots,x_n;\delta;x_i}{M}$,

\item\label{item:strictterms:two}
  $\lfprove{\Gamma,\Gamma_0,\Delta}{\oftype{M}{A}}$, and

\item\label{item:strictterms:three}
  $\lfprove{\Gamma,\subst{\Delta}{N_1/x_1\ldots N_n/x_n}}{\subst{(\oftype{M}{A})}{N_1/x_1\ldots N_n/x_n}}$.
\end{enumerate}
Then there is a derivation of
$\lfprove{\Gamma}{\oftype{N_i}{\subst{B_i}{N_1/x_1...N_{i-1}/x_{i-1}}}}$.
\end{lemma'}
\begin{proof} By induction on the derivation of
$\termUV{x_1,\ldots,x_n;\delta;x_i}{M}$. The argument
proceeds by considering the cases for the last rule used in the
derivation.

\smallskip

\noindent {\it The last rule is \ruvinito}.
In this case, $M$ is $(\app{x_i}{\app{y_1}{\ldots y_k}})$ for some
distinct $y_1,..,y_k\in\delta$ and $y_i$.
Then $\subst{M}{N_1/x_1\ldots N_n/x_n}$ must in fact be
$(\app{N_i}{\app{y_1}{\ldots y_k}})$.
From (\ref{item:strictterms:two}), it follows that $B_i$, the type of
$x_i$, must be $\typedpi{y_1}{C_1}{\ldots\typedpi{y_k}{C_k}{A}}$.
Note that none of the variables in $dom(\Delta)$ can
appear in $B_i$ (and hence $A$) or in $N_j$ for $1 \leq j\leq n$ and,
further, that $B_i$ cannot contain $x_j$ if $j \geq i$.
We then get from (\ref{item:strictterms:three}) that
$\lfprove{\Gamma}{\oftype{(\app{N_i}{\app{y_1}{\ldots y_k}})}
  {\subst{B_i}{N_1/x_1...N_{i-1}/x_{i-1}}}}$
has a derivation. By using below this derivation a sequence of
$\absfam$ rules and using the fact that the variables $y_1, \ldots,
y_k$ cannot appear in $\subst{A}{N_1/x_1...N_{i-1}/x_{i-1}}$, we see then
that there must be a derivation of
\begin{gather*}
\lfprove{\Gamma}{\typedlambda{y_1}{C_1}{\ldots\typedlambda{y_k}{C_k}{(\app{N_i}
      {\app{y_1}{\ldots y_k}})}} : \\
\qquad\qquad\qquad\qquad
\typedpi{y_1}{C_1}{\ldots\typedpi{y_k}{C_k}{\subst{A}{N_1/x_1...N_{i-1}/x_{i-1}}}}}.
\end{gather*}
But $\typedlambda{y_1}{C_1}{\ldots\typedlambda{y_k}{C_k}{(\app{N_i}
    {\app{y_1}{\ldots y_k}})}}$ is equivalent (via the $\eta$-conversion
rule) to $N_i$ and
\[ \typedpi{y_1}{C_1}{\ldots\typedpi{y_k}{C_k}{\subst{A}{N_1/x_1...N_{i-1}/x_{i-1}}}} \]
is identical to
\[ \subst{(\typedpi{y_1}{C_1}{\ldots\typedpi{y_k}{C_k}{A}})}{N_1/x_1...N_{i-1}/x_{i-1}},\]
\ie, to $\subst{B_i}{N_1/x_1...N_{i-1}/x_{i-1}}$, Thus we actually have a
derivation for
\[ \lfprove{\Gamma}{\oftype{N_i}{\subst{B_i}{N_1/x_1...N_{i-1}/x_{i-1}}}}, \]
as desired.

\smallskip

\noindent {\it The last rule is \ruvappo}. In this case, $M=\app{y}{M_1
\ldots M_k}$ for some $y\not\in\Gamma$ of type $\typedpi{z_1}{C_1}{\ldots
\typedpi{z_k}{C_k}{A}}$ and $\termUV{x_1,\ldots,x_n;\delta;x_i}{M_j}$ for
some $j$.
Then by successive applications of \appobj~to (\ref{item:strictterms:two}) there is a
derivation of 
\[ \lfprove{\Gamma,\Gamma_0,\Delta}{\oftype{M_j}{C_j}}. \]
Similarly from (\ref{item:strictterms:three}) we find
\begin{tabbing}
\quad\=\kill
\>$\lfprove{\Gamma,\subst{\Delta}{N_1/x_1\ldots N_n/x_n}}{}$\\
\>\qquad\quad
$\oftype{\subst{M_j}{N_1/x_1\ldots N_n/x_n}}{\subst{C_j}{N_1/x_1\ldots N_n/x_n}}$.
\end{tabbing}
We conclude by using the induction hypothesis.

\smallskip

\noindent {\it The last rule is \ruvabso}.
In this case $M=\typedlambda{y}{C}{M'}$, $A=\typedpi{y}{C}{A'}$ and there is a
derivation of $\termUV{x_1,\ldots,x_n;\delta,y;x_i}{M'}$.
Looking also at the other two derivations, \absobj~provides
derivations of both $$\lfprove{\Gamma,\Gamma_0,\Delta,\oftype{y}{C}}{\oftype{M'}{A'}}$$ and
\begin{gather*}
\lfprove{\Gamma,\subst{\Delta}{N_1/x_1\ldots N_n/x_n},\oftype{y}{\subst{C}{N_1/x_1\ldots
N_n/x_n}}}{\\
 \qquad\qquad\qquad \oftype{\subst{M'}{N_1/x_1\ldots N_n/x_n}}{\subst{A'}{N_1/x_1\ldots N_n/x_n}}}.
\end{gather*}
Now we can conclude by the induction hypothesis.
\end{proof}

\begin{lemma'}\label{lem:stricttypes}
Let $N_1,\ldots,N_n$ be LF objects and $\Gamma$,
$\Gamma_0$, and $\Theta$ be LF contexts with
$\Gamma_0=\oftype{x_1}{B_1},...,\oftype{x_n}{B_n}$.
Further, let 
$\typedpi{x_1}{B_1}{\ldots\typedpi{x_n}{B_n}{A}}$ be an LF type with
$A$ a base type.
Finally, suppose for some $i$ there are derivations of
\begin{enumerate}
\item $\formulaUV{\oftype{x_1}{B_1},\ldots,\oftype{x_n}{B_n},\Theta;x_i}{A}$

\item\label{item:stricttypes:two}
  $\lfprove{\Gamma,\Gamma_0,\Theta}{\oftype{A}{\type}}$ and

\item\label{item:stricttypes:three}
  $\lfprove{\Gamma,\subst{\Theta}{N_1/x_1\ldots
    N_n/x_n}}{\oftype{\subst{A}{N_1/x_1,...,N_n/x_n}}{\type}}$.
\end{enumerate}
Then there is a derivation of
$\lfprove{\Gamma}{\oftype{N_i}{\subst{B}{N_1/x_1...N_{i-1}/x_{i-1}}}}$.
\end{lemma'}
\begin{proof}
We prove the lemma by induction on the structure of the derivation of
$\formulaUV{\oftype{x_1}{B_1},...,\oftype{x_n}{B_n};x_i}{A}$. The argument proceeds by
considering the case for the last rule in the derivation.

\smallskip

\noindent {\it The derivation concludes by \ruvappt.}

Then there is some $c$ of type $\typedpi{y_1}{C_1}{\ldots\typedpi{y_k}{C_k}{A}}$  and $A$
is of the form $\app{c}{M_1\ldots M_k}$ where $x_i$ occurs rigidly in some
$M_j$.
Therefore we have a derivation of
$\termUV{x_1,\ldots,x_n,dom(\Theta);\cdot;x_i}{M_j}$.
Successive applications of \appfam~to (\ref{item:stricttypes:two}) and
  (\ref{item:stricttypes:three})
simultaneously will
provide derivations of $\lfprove{\Gamma,\Gamma_0,\Theta}{\oftype{M_j}{C_j}}$ and
\begin{tabbing}
\quad\=\kill
\>$\lfprove{\Gamma,\subst{\Theta}{N_1/x_1\ldots N_n/x_n}}{}$\\
\>\qquad\quad
$\oftype{\subst{M_j}{N_1/x_1\ldots N_n/x_n}}{\subst{C_j}{N_1/x_1\ldots N_n/x_n}}$.
\end{tabbing}
respectively. We conclude by
invoking Lemma~\ref{lem:strictterms}.

\smallskip

\noindent {\it The derivation concludes by \ruvctxt.}
Then there is some $\oftype{x_j}{B_j}$ such that
\[ \formulaUV{\oftype{x_1}{B_1},\ldots,\oftype{x_{j-1}}{B_{j-1}};x_i}{B_j} \]
 and
\begin{align}
\formulaUV{\oftype{x_1}{B_1},\ldots,\oftype{x_n}{B_n},\Theta;x_j}{A}. \tag{i}\label{eq:one}
\end{align}
Now, by (\ref{eq:one}), (\ref{item:stricttypes:two}) and (\ref{item:stricttypes:three}), there is a derivation of
\begin{center} $\lfprove{\Gamma}{\oftype{N_j}{\subst{B_j}{N_1/x_1\ldots N_{j-1}/x_{j-1}}}}$.\end{center}
And so there must be a derivation of
\begin{align}
\lfprove{\Gamma}{\oftype{\subst{B_j}{N_1/x_1\ldots N_{j-1}/x_{j-1}}}{\type}}. \tag{ii}\label{eq:two}
\end{align}
We also have a derivation of
\begin{align}
\lfprove{\Gamma,\oftype{x_1}{B_1},\ldots,\oftype{x_{j-1}}{B_{j-1}}}{\oftype{B_j}{\type}}. \tag{iii}\label{eq:three}
\end{align}
If $B_j$ is a base type we can conclude by the induction hypothesis.

Otherwise $B_j$ is of the form $\typedpi{y_1}{C_1}{\ldots\typedpi{y_k}{C_k}{D}}$ with $D$
a base type, and by \ruvpit~there is a derivation of
$$\formulaUV{\oftype{x_1}{B_1},\ldots,\oftype{x_{j-1}}{B_{j-1}},\Theta';x_i}{D}$$
with $\Theta'=\oftype{y_1}{C_1},\ldots,\oftype{y_k}{C_k}$.
From (\ref{eq:two}) and (\ref{eq:three}) we obtain derivations of
$$\lfprove{\Gamma,\subst{\Theta'}{N_1/x_1\ldots
N_{j-1}/x_{j-1}}}{\oftype{\subst{D}{N_1/x_1\ldots N_{j-1}/x_{j-1}}}{\type}}$$
and $$\lfprove{\Gamma,\Theta'}{\oftype{D}{\type}}$$ by \pifam.
Since $D$ must be a base type we can conclude by the induction hypothesis.
\end{proof}

From these lemmas we can conclude that explicitly checking that the types of
strict variables are inhabited is redundant as this is already
guaranteed by the formation of the target type.
We now to turn leveraging this to improve the \hhf~clauses generated
by translation of LF signatures.

\subsection{Eliminating Redundancies in the Translation}\label{sec:refined:trans}
In the simple translation presented in Section~\ref{sec:naive}, every
binder will generate a $hastype$ formula which amounts to showing that
a term is an inhabitant of a type.
But we have shown that substitutions for strict variables must inhabit
the correct type, and we would like to modify the translation so that
this is taken into account.
There are now two modes in which translation operates, the negative,
$\encExtN{\cdot}$, which is essentially the same as before in that it
does not check for strictness of bound variables, and the positive,
$\encExtP{\cdot}{}$, which will only generate $hastype$ formulas for
variables which do not appear strictly.
We do this to insure that the eliminations occur in situations in
which it makes sense to think of the implication encoding an
inhabitation check.
The new rules for translating judgments can be found in
Figure~\ref{fig:optimized-translation}.
The encoding of types and objects remains the same as in the simple
translation.

\begin{figure*}
  \centering
  \begin{align*}
    \encExtP{\typedpi{x}{A}{B}}{\Gamma} :=&\
      \begin{cases}
        \lambda M.~ \forall x.~ \top \supset \encExtP{B}{\Gamma, x}(\app{M}{x})
          & \text{if}\ \formulaUV{\Gamma; x}{B} \\
        \lambda M.~ \forall x.~
            \encExtN{A}(x) \supset \encExtP{B}{\Gamma, x}(\app{M}{x})
          & \text{otherwise}
      \end{cases} \\
    \encExtP{u \vec{N}}{\Gamma} :=&\ \lambda M.~ \hastype{M}{(\app{u}{\vec{\encTerm{N}}})}
    \\
    \medskip
    \\
    \encExtN{\typedpi{x}{A}{B}} :=&\
      \lambda M.~ \forall x.~
         \encExtP{A}{\cdot}(x) \supset \encExtN{B}(M \app x) \\
    \encExtN{u \vec{N}} :=&\ \lambda M.~ \hastype{M}{(\app{u}{\vec{\encTerm{N}}})}
  \end{align*}
\nocaptionrule  \caption{Optimized translation of LF specifications and judgments to \hhf}
  \label{fig:optimized-translation}
\end{figure*}

%
It is useful to define a notion of equivalence between two \hhf~terms which
encode the same LF term.
The idea is that these \hhf~terms are equivalent up to some
substitution by terms $t_1,\ldots,t_n$.
We say $\subst{(\encEq{M'}{M})}{t_1/x_1\ldots t_n/x_n}$
when $$\encTerm{M'}=\subst{\encTerm{M}}{t_1/x_1\ldots t_n/x_n}.$$
We can use the same notation to give some idea of equality between
types, where $\subst{(\encEq{A'}{A})}{t_1/x_1\ldots t_n/x_n}$ when the
term equality holds on all objects in the type.
Lastly, the notion of
$\subst{(\encEq{\Gamma'}{\Gamma})}{t_1/x_1\ldots t_n/x_n}$ for contexts
just pushes the equality down to each type bound in the context.
In general, the substitution $\subst{}{t_1/x_1\ldots t_n/x_n}$ is
omitted when it is clear which term substitutions are being used.

The lemmas presented above show a desirable property of strict
variables in the LF setting.
It is of interest to consider if they still hold under translation
to \hhf~clauses.
In effect, we would like to show that for LF types $A$ and $A'$ and \hhf~terms
$t_1,\ldots,t_n$ such that $\subst{(\encEq{A'}{A})}{t_1/x_1\ldots t_n/x_n}$,
if $x_i$ appears strictly in $A$ then $t_i$ is the encoding of some LF
term $t_i'$.
The following lemmas formalize this idea and the proofs follow the
structure of the derivation of strictness.

\begin{lemma'}\label{lem:lfterms}
Let $\Gamma$, $\Delta_1$ and $\Delta_2$ be valid LF contexts, and let
$\delta\subseteq dom(\Delta_1)$.
Suppose there are derivations of
\begin{enumerate}
\item\label{item:terms:one}
$\lfprove{\Gamma,\oftype{x_1}{B_1},\ldots,\oftype{x_n}{B_n},\Delta_1}{\oftype{M}{A}}$

\item\label{item:terms:two}
$\termUV{x_1,\ldots,x_n;\delta;x_i}{M}$, and

\item\label{item:terms:three}
$\lfprove{\Gamma,\Delta_2}{\oftype{M'}{A'}}$
\end{enumerate}
where $\subst{(\encEq{M'}{M})}{t_1/x_1,\ldots,t_n/x_n}$ for some
\hhf~terms $t_1,\ldots,t_n$.
Then there is an LF term $t_i'$ such that $t_i=\encTerm{t_i'}$.
\end{lemma'}
\begin{proof}
The proof proceeds by induction on the structure of the derivation 
$\termUV{x_1,\ldots,x_n;\delta;x_i}{M}.$

\smallskip

\noindent{\it The derivation concludes by \ruvinito.}
Then $M$ has the form $\app{x_i}{y_1\ldots y_k}$ for distinct $y$ from $\delta$.
So $t_i$ must be a term of the form $$\lambdax{z_1}{\ldots\lambda{z_k}{u}}.$$
We use $\encEq{M'}{M}$ to determine that $\encTerm{M'}=\subst{u}{y_1/z_1\ldots y_k/z_k}$.
\begin{align*}
\encTerm{M'} &=\subst{\encTerm{M}}{t_1/x_1\ldots t_i/x_i}\\
             &=\subst{\encTerm{\app{x_i}{y_1\ldots y_k}}}{t_1/x_1\ldots t_i/x_i}\\
             &=\subst{(\app{x_i}{y_1\ldots y_k})}{t_1/x_1\ldots t_i/x_i}\\
             &= \app{t_i}{y_1\ldots y_k}\\
             &= \subst{u}{y_1/z_1\ldots y_k/z_k}
\end{align*}
Because the $y_i$'s are distinct the substitution on $u$ can be inverted
and we find $u=\subst{\encTerm{M'}}{y_1/z_1\ldots y_k/z_k}$.
In fact, because the encoding leaves variables unchanged we can
determine that 
$\subst{\encTerm{M'}}{y_1/z_1\ldots y_k/z_k}=\encTerm{\subst{M'}{y_1/z_1\ldots y_k/z_k}}$.
Each $y_j$ is a distinct variable from $\delta$,
and so for some type $C_j$, there is a binding $\oftype{y_j}{C_j}\in\Delta_1$.
Let $t_i'=\typedlambda{z_1}{C_1}{\ldots\typedlambda{z_k}{C_k}{u'}}$
where $u'=\subst{M'}{y_1/z_1\ldots y_k/z_k}$.
Then $\encTerm{t_i'}=t_i$.
\begin{align*}
\encTerm{t_i'} &= \encTerm{\typedlambda{z_1}{C_1}{\ldots\typedlambda{z_k}{C_k}{u'}}}\\
               &= \lambdax{z_1}{\ldots\lambdax{z_k}{\encTerm{\subst{M'}{y_1/z_1\ldots y_k/z_k}}}}\\
               &= \lambdax{z_1}{\ldots\lambdax{z_k}{\subst{\encTerm{M'}}{y_1/z_1\ldots y_k/z_k}}}\\
               &= \lambdax{z_1}{\ldots\lambdax{z_k}{u}}\\
               &= t_i
\end{align*}

\smallskip

\noindent{\it The derivation concludes by \ruvabso.}
Then $M$ is of the form $\typedlambda{y}{C}{N}$ and from (\ref{item:terms:two})
we have a derivation of $\termUV{x_1,\ldots,x_n;\delta,y;x_i}{N}$.
Because $\encEq{M'}{M}$ their structures must be similar,
and so $M'$ has the form $\typedlambda{y}{C'}{N'}$ with $\encEq{N'}{N}$.
The type $A$ must be of the form $\typedpi{y}{C}{D}$, and $A'$ will be of
the form $\typedpi{y}{C'}{D'}$.
From (\ref{item:terms:one}) and (\ref{item:terms:three}) respectively,
\absobj~provides derivations of
$$\lfprove{\Gamma,\oftype{x_1}{B_1},\ldots,\oftype{x_n}{B_n},\Delta_1,\oftype{y}{C}}{\oftype{N}{C}}$$
and $\lfprove{\Gamma,\Delta_2,\oftype{y}{C'}}{\oftype{N'}{C'}}.$
We can conclude by the inductive hypothesis.

\smallskip

\noindent{\it The derivation concludes by \ruvappo.}
Then $M$ has the form $\app{y}{N_1\ldots N_k}$ with $y\not=x_j$ for $j\leq i$.
So by the rules of strictness, there is a derivation of
$\termUV{x_1,\ldots,x_n;\delta;x_i}{N_l}$ for some $l<k$.
Successive applications of \appobj~to (\ref{item:terms:one}) followed by \varobj~on some
$\oftype{y}{\typedpi{z_1}{C_1}{\ldots\typedpi{z_k}{C_k}{D}}}$
provide a derivation of
$$\lfprove{\Gamma,\oftype{x_1}{B_1},\ldots,\oftype{x_n}{B_n},\Delta_1}{\oftype{N_l}{\subst{C_l}{N_1/z_1\ldots N_{l-1}/z_{l-1}}}}.$$
Since $\encEq{M'}{M}=\app{y}{N_1\ldots N_k}$, $M'$ has the form $\app{y}{N_1'\ldots N_k'}$ where $\encEq{N_l'}{N_l}$.
It should be clear that the derivation of (\ref{item:terms:three})
proceeded similarly to (\ref{item:terms:one}) but with some
$\oftype{y'}{\typedpi{z_1'}{C_1'}{\ldots\typedpi{z_k'}{C_k'}{D'}}}$.
Thus we find a derivation of
$$\lfprove{\Gamma,\Delta_2}{\oftype{N_l'}{\subst{C_l'}{N_1'/z_1'\ldots N_{l-1}'/z_{l-1}'}}}.$$
We conclude by the induction hypothesis.
\end{proof}

\begin{lemma'}\label{lem:lftypes}
Let $\Gamma$, $\Theta_1$, and $\Theta_2$ be valid LF contexts and
\[ \typedpi{x_1}{B_1}{\ldots\typedpi{x_n}{B_n}{A}} \] 
be a type with
$A$ a base type.
Suppose there are derivations of
\begin{enumerate}
\item\label{item:types:one}
$\lfprove{\Gamma,\oftype{x_1}{B_1},\ldots,\oftype{x_n}{B_n},\Theta_1}{\oftype{A}{\type}}$

\item\label{item:types:two}
$\formulaUV{\oftype{x_1}{B_1},\ldots,\oftype{x_n}{B_n};x_i}{A}$, and

\item\label{item:types:three}
$\lfprove{\Gamma,\Theta_2}{\oftype{A'}{\type}}$
\end{enumerate}
where $\subst{(\encEq{A'}{A})}{t_1/x_1,\ldots,t_n/x_n}$ for
\hhf~terms $t_1,\ldots,t_n$.
Then there is an LF term $t_i'$ such that $t_i=\encTerm{t_i'}$.
\end{lemma'}
\begin{proof}
Proceed by induction on the derivation of
$$\formulaUV{\oftype{x_1}{B_1},\ldots,\oftype{x_n}{B_n};x_i}{A}.$$
Note that the rule \ruvpit~will not apply because we know $A$ to be a base type.
\smallskip

\noindent{\it The derivation concludes by \ruvappt.}
Then $A$ is a base type of the form
$\app{c}{N_1\ldots N_k}$ and there is a derivation of
$\formulaUV{x_1,\ldots,x_n;\cdot;x_i}{N_l}$ for some $l<k$.
Applying \appfam~to (\ref{item:types:one}) provides the derivation of
$$\lfprove{\Gamma,\oftype{x_1}{B_1},\ldots,\oftype{x_n}{B_n},\Theta_1}{\oftype{N_l}{T}}.$$
Because $\encEq{A'}{A}$, their structures must be similar and so
$A'$ has the form $\app{c}{N_1'\ldots N_k'}$ with $\encEq{N_j'}{N_j}$.
Thus from (\ref{item:types:three}) we can obtain a derivation of
$\lfprove{\Gamma,\Theta_2}{\oftype{N_l'}{T'}}$
where $\encEq{N_1'}{N_l}$.
We conclude by Lemma \ref{lem:lfterms}.

\smallskip

\noindent{\it The derivation concludes by \ruvctxt.}
Then there must be derivations of
\begin{align*}
\formulaUV{\oftype{x_1}{B_1},\ldots,\oftype{x_{j-1}}{B_{j-1}};x_i}{B_j}\tag{i}\label{eq:strictctx:one}
\end{align*}
and $\formulaUV{\oftype{x_1}{B_1},\ldots,\oftype{x_n}{B_n};x_j}{A}.$
From the assumption (\ref{item:types:two}) there is a derivation of
$\lfprove{\Gamma,\oftype{x_1:B_1},\ldots,\oftype{x_{j-1}}{B_{j-1}}}{\oftype{x_j}{B_j}}$
and so also one of
$$\lfprove{\Gamma,\oftype{x_1:B_1},\ldots,\oftype{x_{j-1}}{B_{j-1}}}{\oftype{B_j}{\type}}.$$
By the induction hypothesis on (\ref{eq:strictctx:one}), there is an LF object
$t_j'$ such that $t_j=\encTerm{t_j'}$.
In addition, $t_j'$ must be of some type $B_j'$ where $\encEq{B_j'}{\subst{B_j}{t_1/x_1\ldots t_n/x_n}}$.
Thus we conclude by the induction hypothesis.
\end{proof}

We can now assume that any substitution $t$ for a strict variable $x$
which is performed in the translation, in fact is the encoding of a
valid LF term $t'$.
Because $t=\encTerm{t'}$ we can use Lemma~\ref{lem:stricttypes} to guarantee
$t'$ inhabits the correct type.
Thus it is reasonable to remove the \hhf~formula which causes us to
explicitly show that $\hastype{t}{A}$ holds for the appropriate type.

We show this new translation to be correct by relating the new
translation with the simple one given in Section~\ref{sec:naive}.
From the correctness of this simple
translation (Theorem~\ref{thm:naive}) then, we get that the
improved translation is also correct.
Showing completeness is straightforward because it simply erases information from
derivations resulting from the simple translation.
Soundness is more involved.
We must reconstruct a derivation which has been eliminated by the
strictness condition using properties of strict variables which have
been shown through Lemmas~\ref{lem:stricttypes} and~\ref{lem:lftypes}.

\begin{theorem'}\label{thm:improvedcorrectness}
Let $\Gamma$ be a valid LF context and $A$ an LF type such that 
\begin{enumerate}
\item\label{item:typea}$\lfprove{\Gamma}{\oftype{A}{\type}}$
\end{enumerate}
is derivable.
Then for arbitrary \hhf~term $M$, $\iprove{\enc{\Gamma}}{\enc{A}(M)}$
has a derivation if and only if
$\iprove{\encExtP{\Gamma}{}}{\encExtN{A}(M)}$
has a derivation.
\end{theorem'}
\begin{proof}

\smallskip

\noindent{\it Completeness}

We proceed by induction on the derivation of $\iprove{\enc{\Gamma}}{\enc{A}(M)}.$

\smallskip

\noindent{\it The derivation concluded by \allGoal}.
Then $A$ must have the form $\typedpi{x}{B}{A'}$, and
 the derivation concluded by \allGoal~and \impGoal~as shown below.
\begin{center}
\AxiomC{$\iprove{\enc{\Gamma,\oftype{x}{B}}}{\enc{A'}(\app{M}{x})}$}
\RightLabel{\allGoal, \impGoal}
\UnaryInfC{$\iprove{\enc{\Gamma}}{\app{\enc{\typedpi{x}{B}{A'}}}{(M)}}$}
\DisplayProof
\end{center}
By assumption (\ref{item:typea}), and the derivation rules of LF,
there must be a derivation of
$\lfprove{\Gamma,\oftype{x}{B}}{\oftype{A'}{\type}}.$
By the induction hypothesis there is then a derivation of
$\iprove{\encExtP{\Gamma,\oftype{x}{B}}{}}{\encExtN{A'}(\app{M}{x})}.$
From this we can construct the following:
\begin{center}
\AxiomC{$\iprove{\encExtP{\Gamma,\oftype{x}{B}}{}}{\encExtN{A'}(\app{M}{x})}$}
\RightLabel{\allGoal, \impGoal}
\UnaryInfC{$\iprove{\encExtP{\Gamma}{}}{\encExtN{A}(M)}$}
\DisplayProof
\end{center}
which is exactly the derivation desired.

\smallskip

\noindent{\it The derivation concludes by \impGoal.}
This case proceeds as for the previous case.

\smallskip

\noindent{\it The derivation concludes by \bcGoal~on the encoding of a term \\
$(\oftype{y}{\typedpi{x_1}{B_1}{\ldots\typedpi{x_n}{B_n}{A'}}})\in\Gamma$.}
Then $A$ is a base type of the form $\app{u}{N_1\ldots N_k}$.
It is clear then that $M$ is of the form $\app{y}{t_1\ldots t_n}$ for some \hhf~terms $t_1,\ldots t_n$
and $\subst{\encEq{A}{A'}}{t_1/x_1\ldots t_n/x_n}$.
We are left with the collection of derivations shown below.
\begin{center}
\AxiomC{$\{\iprove{\enc{\Gamma}}{\subst{(\enc{B_i} x_i)}{t_1/x_1\ldots t_{i-1}/x_{i-1}}}\}_{0<i\leq n}$}
\RightLabel{\bcGoal}
\UnaryInfC{$\iprove{\enc{\Gamma}}{\hastype{(\app{y}{t_1\ldots t_n})}{(\app{u}{\encTerm{N_1}}\ldots\encTerm{N_k})}}$}
\DisplayProof
\end{center}
To continue we show an inner induction on $i$ asserting that if for
 $j< i$ $t_j=\encTerm{t_j'}$ for some LF object $t_j'$,
 then there is a $t_i'$ such that $t_i=\encTerm{t_i'}$.\\
Because $t_j=\encTerm{t_j'}$ for $j<i$, we can move the
substitution inside of the encoding giving the equivalence
$$\subst{(\app{\enc{B_i}}{x_i})}{t_1/x_1\ldots t_{i-1}/x_{i-1}} = \app{\enc{\subst{B_i}{t_1'/x_1'\ldots t_{i-1}'/x_{i-1}}}}{t_i}.$$
So there is a derivation
$\iprove{\enc{\Gamma}}{\app{\enc{\subst{B_i}{t_1'/x_1'\ldots t_{i-1}'/x_{i-1}}}}{t_i}}.$
We conclude by the correctness of the naive translation that
there is an LF object $t_i'$ such that $t_i=\encTerm{t_i'}$.\\
By our assumption, $\Gamma$ is a valid context and so
$$\lfprove{\Gamma}{\oftype{\typedpi{x_1}{B_1}{\ldots\typedpi{x_n}{B_n}{A'}}}{\type}}$$
is derivable.
We substitute the $t_i'$'s into this derivation to construct one of
$\lfprove{\Gamma}{\oftype{\subst{B_i}{t_1'/x_1\ldots t_{i-1}'/x_{i-1}}}{\type}}$
for each $i$.
By the outer induction we obtain the collection of derivations
$$\{\iprove{\encExtP{\Gamma}{}}{\app{\encExtN{\subst{B_i}{t_1'/x_1\ldots t_{i-1}'/x_{i-1}}}}{t_i}}\}_{0<i\leq n}.$$
To conclude we apply \bcGoal~with the improved translation of the same term
$(\oftype{y}{\typedpi{x_1}{B_i}{\ldots\typedpi{x_n}{B_n}{A'}}})\in\Gamma$ selecting the derivation
$\iprove{\encExtP{\Gamma}{}}{\top}$ if there is a strict occurrence of $x_i$ and
$$\iprove{\encExtP{\Gamma}{}}{\app{\encExtN{\subst{B_i}{t_1'/x_1\ldots t_{i-1}'/x_{i-1}}}}{t_i}}$$ otherwise.
Thus we have a derivation of
$$\iprove{\encExtP{\Gamma}{}}
{\hastype{(\app{y}{t_1\ldots t_n})}{\subst{(\app{u}{\encTerm{N_1}\ldots\encTerm{N_1}})}{t_1/x_1\ldots t_n/x_n}}}.$$
as desired.

\smallskip

\noindent{\it Soundness}

We proceed by induction on the derivation of $\iprove{\encExtP{\Gamma}{}}{\encExtN{A}(M)}.$

\smallskip

\noindent{\it The derivation concludes by \allGoal.}
Then $A$ must be of the form $\typedpi{x}{B}{A'}$,
and the derivation concludes by both \allGoal~and \impGoal~as shown below.
\begin{center}
\AxiomC{$\iprove{\encExtP{\Gamma,\oftype{x}{B}}{}}{\app{\encExtN{A'}}{(\app{M}{x})}}$}
\RightLabel{\allGoal, \impGoal}
\UnaryInfC{$\iprove{\encExtP{\Gamma}{}}{\app{\encExtN{\typedpi{x}{B}{A'}}}{(M)}}$}
\DisplayProof
\end{center}
By assumption (\ref{item:typea}) there are derivations of
$\lfprove{\Gamma}{\oftype{B}{\type}}$ and $\lfprove{\Gamma,\oftype{x}{B}}{\oftype{A'}{\type}}.$
From this and the assumption $\Gamma$ is a valid context, we can construct a
derivation of $\lfprove{}{\Gamma,\oftype{x}{B}\ctx}.$
Now by the inductive hypothesis there is a derivation of
$$\iprove{\enc{\Gamma,\oftype{x}{B}}}{\enc{A'}(\app{M}{x})}.$$
To conclude we build the desired derivation as shown below.
\begin{center}
\AxiomC{$\iprove{\enc{\Gamma,\oftype{x}{B}}}{\enc{A'}(\app{M}{x})}$}
\RightLabel{\allGoal, \impGoal}
\UnaryInfC{$\iprove{\enc{\Gamma}}{\enc{A}(M)}$}
\DisplayProof
\end{center}

\smallskip

\noindent{\it The derivation concludes by \impGoal.}
This case follows that for the previous case.

\smallskip

\noindent{\it The derivation concludes by \bcGoal~on the encoding of a term \\
$\oftype{y}{\typedpi{x_1}{B_1}{\ldots\typedpi{x_n}{B_n}{A'}}}\in\Gamma$.}
Then $A$ must be a base type.
This means $M$ is of the form $\app{y}{t_1\ldots t_n}$ for some \hhf~terms $t_1,\ldots, t_n$,
and we have $\subst{\encEq{A}{A'}}{t_1/x_1\ldots t_n/x_n}$.
Because $\Gamma$ is a valid context,
$\lfprove{\Gamma,\oftype{x_1}{B_1},\ldots,\oftype{x_n}{B_n}}{\oftype{A'}{\type}}$ has a derivation.
Then by Lemma \ref{lem:lftypes}, if $x_i$ has a strict occurrence in $A'$
there must be an LF term $t_i'$ such that $t_i=\encTerm{t_i'}$.
Since $A'$ must then also be a base type, let it be of the form $\app{u}{N_1\ldots N_k}$.
Then by the definition of \bcGoal~there are a collection of derivations
$\{\iprove{\encExtP{\Gamma}{}}{F_i}\}_{i}$
where $F_i=\top$ if $x_i$ appears strictly and
$$F_i=\subst{(\app{\encExtN{B_i}}{(x_i)})}{t_1/x_1\ldots t_n/x_n}$$ otherwise.\\
We continue by showing an inner induction on $i$ that
if for each $j<i$, $t_j=\encTerm{t_j'}$ for some LF object $t_j'$
then there is $t_i'$ such that $t_i=\encTerm{t_i'}$.

\smallskip

\noindent{\it Suppose $F_i=\top$.}\\
Then it must be that $x_i$ appears rigidly
and such a $t_i'$ exists by the argument stated previously.

\smallskip

\noindent{\it Suppose $F_i=\subst{(\encExtN{B_i}(x_i))}{t_1/x_1\ldots t_i/x_i})$.}\\
Since $\Gamma$ is a valid context,
$\lfprove{\Gamma}{\oftype{\typedpi{x_1}{B_1}{\ldots\typedpi{x_n}{B_n}{A'}}}{\type}}$
must be derivable.
Then for each $i$, there must also be a derivation of
$\lfprove{\Gamma,\oftype{x_1}{B_1},\ldots,\oftype{x_{i-1}}{B_{i-1}}}{\oftype{B_i}{\type}}.$
Because $t_j=\encTerm{t_j'}$ for each $t_j$, we can substitute
these LF objects to obtain a derivation of
$\lfprove{\Gamma}{\oftype{\subst{B_i}{t_1'/x_1\ldots t_{i-1}'/x_{i-1}}}{\type}},$
and so by the outer induction, there must be a derivation of
$$\iprove{\enc{\Gamma}}{\app{\enc{\subst{B_i}{t_1'/x_1\ldots t_{i-1}'/x_{i-1}}}}{t_i}}.$$
By the correctness of the naive translation, $t_i=\encTerm{t_i'}$
for some LF object $t_i'$.

\smallskip

We now construct a derivation of
$$\iprove{\enc{\Gamma}}{\app{\enc{\subst{B_i}{t_1'/x_1\ldots t_{i-1}'/x_{i-1}}}}{t_i}}$$
for each $i$.

\smallskip

\noindent{\it Suppose $F_i=\subst{(\encExtN{B_i}(x_i))}{t_1/x_1\ldots t_i/x_i})$.}\\

Then the derivation was found in the previous argument.

\smallskip

\noindent{\it Suppose $F_i=\top$}\\

Then there is a derivation of $\formulaUV{x_1,\ldots,x_n;x_i}{A'}.$
We also have a derivation of
$\lfprove{\Gamma}{\oftype{\typedpi{x_1}{B_1}{\ldots\typedpi{x_n}{B_n}{A'}}}{\type}}.$
Both $A$ and $A'$ are base types, and so since $\encEq{A}{A'}$, we know that
$\encTerm{A}=\subst{\encTerm{A'}}{t_1/x_1\ldots t_n/x_n}$.
We have shown that for each $i$, $t_i=\encTerm{t_i'}$ for an LF object $t_i'$.
And thus $$\subst{\encTerm{A'}}{t_1/x_1\ldots t_n/x_n}=\encTerm{\subst{A'}{t_1'/x_1\ldots t_n'/x_n}}.$$
By injectivity of $\encTerm{\cdot}$ then,
$A=\subst{A'}{t_1'/x_1\ldots t_n'/x_n}$
and by assumption (\ref{item:typea}) we have a derivation of
$\lfprove{\Gamma}{\subst{A'}{t_1'/x_1\ldots t_n'/x_n}}.$
Applying Lemma \ref{lem:stricttypes} we now obtain a derivation of
$$\lfprove{\Gamma}{\oftype{t_i'}{\subst{B_i}{t_1'/x_1\ldots t_{i-1}'/x_{i-1}}}}.$$
Finally, the correctness of the naive translation allows us to conclude there is a derivation of
$\iprove{\enc{\Gamma}}{\app{\enc{\subst{B_i}{t_1'/x_1\ldots t_n'/x_n}}}{t_i}}$
as desired.

Now we have a collection of derivations which we compose using
\bcGoal~to obtain a derivation of
$$\iprove{\enc{\Gamma}}{\app{\enc{A}}{(\app{y}{t_1\cdots t_n})}}.$$
\end{proof}

The key to this proof is that we are able to reconstruct typing
derivations for instantiations of strict variables by leveraging our
assumption that $A$ is a well formed type in the context $\Gamma$.
By removing the redundant formulas from implications we simplify the
translated signature in a way that more closely resembles the original
LF types.
Recall the signature for append presented
in Figure~\ref{fig:rel-ex} and its translation into \hhf~shown in
Figure~\ref{fig:append-translation}.
Using this new translation, we refine the clauses generated
for $appNil$ and $appCons$, as seen in Figure~\ref{fig:app-improved}.
The clauses shown in the figure are presented in a simplified form where the obviously
satisfiable goals generated by the translation are removed for clarity.
\begin{figure}
  \begin{tabbing}
  \=\qquad\=\qquad\qquad\=\kill
  \> $\hastype{z}{nat}$ \\
  \> $\forall n.~ \hastype{n}{nat} \supset \hastype{(s~n)}{nat}$\\
  \> $\hastype{nil}{list}$ \\
  \> $\forall n.~ \hastype{n}{nat} \supset \forall l.~ \hastype{l}{list} \supset$\\
  \>\> $\hastype{(cons~n~l)}{list}$ \\
  \\
  \> $\forall l.~\hastype{(appNil~l)}{(append~nil~l~l)}$ \\
  \> $\forall x.\forall l.\forall k.\forall m.\forall a.~ \hastype{a}{(append~l~k~m)} \supset$\\
  \>\>$hastype~(appCons~x~l~k~m~a)$\\
  \>\>\>$(append~(cons~x~l)~k~(cons~x~m))$
  \end{tabbing}

\nocaptionrule  \caption{Improved translation of $append$}
  \label{fig:app-improved}
\end{figure}

The result of this example is exactly that which is produced by the
translation based on the former notion of strictness in
\cite{snow10ppdp}, since every strict variable actually
appears in the target type.
Instead consider the example deceleration for $f$ from Figure~\ref{fig:strict-ex}.
Under the previous notion of strictness this would become the
\hhf~clause
\begin{tabbing}
\=\qquad\=\quad\=\kill 
  \>\>$\forall x.(\forall x_1.(\hastype{x_1}{nat}\supset\hastype{(x~x_1)}{nat})\supset$\\
  \>\>\>$\forall y.(\hastype{(f~x~y)}{(d~(y\backslash~y)~(w\backslash y\backslash~x~(w~y))~y)})).$
\end{tabbing}
Using the extension to strictness presented in this work, we can
further reduce this clause by removing the formula related to typing
$x$ resulting in the clause
\[ \forall x.(\forall y.(\hastype{(f~x~y)}{(d~(y\backslash~y)~(w\backslash y\backslash~x~(w~y))~y)})). \]

In Lemma~\ref{lem:lftypes} we are able to take the term $t$, for which
we have some knowledge of the structure, and construct an LF term $t'$
such that $t=\encTerm{t'}$.
In next section we look at how we might encode general \hhf~terms, for
which some type information is known, back into LF.

  \section{Translating back to LF terms}
\label{sec:inverse}
\begin{figure*}
  \centering

  \AxiomC{$\invprove{\Gamma,x:A}{\Sigma}{M}{M'}{B}$}
  \RightLabel{\invabs}
  \UnaryInfC{$\invprove{\Gamma}{\Sigma}{(\lambdax{x}{M})}{(\typedlambda{x}{A}{M'})}{\typedpi{x}{A}{B}}$}
  \DisplayProof\\
\medskip
  \AxiomC{$c:\typedpis{x}{T}{A}\in\Sigma \qquad
           \invprove{\Gamma}{\Sigma}{M_1}{M_1'}{T_1} \qquad \ldots \qquad
           \invprove{\Gamma}{\Sigma}{M_i}{M_i'}{\subst{T_i}{M_1'/x_1\ldots M_{i-1}'/x_{i-1}}}$}
  \RightLabel{\invconst}
  \UnaryInfC{$\invprove{\Gamma}{\Sigma}{(\app{c}{M_1\ldots M_k})}{(\app{c}{M_1'\ldots M_k'})}{A}$}
  \DisplayProof\\
\medskip
  \AxiomC{$x:\typedpis{y}{T}{A}\in\Gamma \qquad
           \invprove{\Gamma}{\Sigma}{M_1}{M_1'}{T_1} \qquad \ldots \qquad
           \invprove{\Gamma}{\Sigma}{M_i}{M_i'}{\subst{T_i}{M_1'/y_1\ldots M_{i-1}'/y_{i-1}}}$}
  \RightLabel{\invvar}
  \UnaryInfC{$\invprove{\Gamma}{\Sigma}{(\app{x}{M_1\ldots M_k})}{(\app{x}{M_1'\ldots M_k'})}{A}$}
  \DisplayProof
\nocaptionrule  \caption{Translating \hhf~terms back to LF}
  \label{fig:reverse-encoding}
\end{figure*}

The previous sections have defined a translation which provides a
means for taking LF specifications to \lprolog~programs.
We would like to use this translation as a vehicle for
efficiently executing LF specifications, but to do this we need a
method for getting back to LF expressions once the execution in
the \hhf~setting is completed.
As we have presented it up to this point, the main objective of proof
search is to identify an LF object corresponding to a given type.
Thus, to make this approach to implementing Twelf work, what we need
is a way to map an \hhf~term back into an LF object.

From the term encoding rules in figure~\ref{fig:naive-translation}
it is clear that through type erasure it is possible for a single
\hhf~term to be the encoding of multiple objects in LF.
This lack of uniqueness poses a problem for defining a general reverse
encoding into LF.
However, if the LF typing information is retained there is in fact a
unique LF expression that the \hhf~term can be identified with.
In the implementation context we are looking at, the relevant typing
information is available directly from the original Twelf query.
Thus after execution, the type can be used in guiding a reverse encoding.

To spell out the inverse translation process more precisely, we will
assume that it takes place in a setting where the type of the LF
object that is to be produced is known.
Moreover, we will assume that each \hhf~constant corresponds to an
object level constant with the same name and a known type; the type
information will be provided by a signature $\Sigma$.
The \hhf~term may contain free variables and we will also assume
that the LF types corresponding to these variables are given by a
context $\Gamma$.
In the beginning, there will be no free variables so
$\Gamma$ can be empty and and our rules will ensure that any time a
free variable is introduced its type is known and so can be added to
$\Gamma$.
Figure~\ref{fig:reverse-encoding} presents rules for deriving
judgments of the form $\invprove{\Gamma}{\Sigma}{M}{M'}{T}$ where $M$
is an \hhf~term and $M'$ is an LF object.
These rules are intended to
be used with $M'$ being the only unknown that is in fact to be
reconstructed by rule applications.
If such an LF expression can be found, then we say that
$M'=\encInv{M}{T}$.
The structure of the (normal-form) term $M$ drives this reverse
encoding.
When an abstraction is encountered, the type of the LF object to be
extracted must obviously be of the form $\typedpi{x}{A}{B}$.
From this knowledge we can easily extract the correct type for the
abstracted variable.
Applications have a simple recursive structure which include a check
that a constant (or variable) of the appropriate type exists in the
signature (context).

Now we must show that this encoding is sensible, in that it will
return meaningful LF expressions in relation to the LF specification.
It must be shown that $\encInv{\cdot}{\tau}$ is in some sense an inverse of
$\encTerm{\cdot}$ on LF terms.
This is done by proving that the composition of these two encodings is
the identity.
Because they are both defined based on the structure of a term $M$,
the structure of this term is used to drive the inductive proof.

\begin{theorem'}
Suppose $M$ is an LF term such that $\Gamma\vdash_{\Sigma}\oftype{M}{T}$ is derivable.
Then $\encInv{\encTerm{M}}{T}=M$.
\ie~there is a derivation of $\invprove{\Gamma}{\Sigma}{\encTerm{M}}{M}{T}$.
\end{theorem'}
\begin{proof}

\smallskip

The proof proceeds by induction on the structure of $M$.

\smallskip

\noindent{\it The term is of the form $\typedlambda{x}{A}{M'}$.}

Then the type $T$ has the form $\typedpi{x}{A}{B}$ and there is a
derivation of $\Gamma,x:A\vdash_{\Sigma}\oftype{M'}{B}.$
So by the induction hypothesis, $\encInv{\encTerm{M'}}{B}=M'$
\ie~there is a derivation of $$\invprove{\Gamma,x:A}{\Sigma}{\encTerm{M'}}{M'}{B}.$$
This and \invabs~provide a derivation of
$$\invprove{\Gamma}{\Sigma}{(\lambdax{x}{\encTerm{M'}})}{(\typedlambda{x}{A}{M'})}{\typedpi{x}{A}{B}}.$$
By definition of $\encTerm{\cdot}$, $\encTerm{M}=\lambdax{x}{\encTerm{M'}}$,
and so we actually have a derivation of
$$\invprove{\Gamma}{\Sigma}{\encTerm{\typedlambda{x}{A}{M'}}}{(\typedlambda{x}{A}{M'})}{\typedpi{x}{A}{B}}$$
as desired.

\smallskip

\noindent{\it The term is of the form $\app{c}{M_1\ldots M_n}$ for some constant $c$.}
Then it must be that $\oftype{c}{\typedpi{x_1}{T_1}{\ldots\typedpi{x_n}{T_n}{T}}}\in\Sigma$.
From this we can be sure there are derivations of
$$\Gamma\vdash_{\Sigma}\oftype{M_i}{\subst{T_i}{M_1/x_1\ldots M_{i-1}/x_{i-1}}}$$
for each $i$.
So by the induction hypothesis 
$$\invprove{\Gamma}{\Sigma}{\encTerm{M_i}}{M_i}{\subst{T}{M_1/x_1\ldots M_{i-1}/x_{i-1}}}$$
is derivable for each $i$.
We can now compose these by \invconst~for a derivation 
$$\invprove{\Gamma}{\Sigma}{(\app{c}{\encTerm{M_1}\ldots\encTerm{M_n}})}{(\app{c}{M_1\ldots M_n})}{T}.$$
But $\encTerm{\app{c}{M_1\ldots M_n}}=(\app{c}{\encTerm{M_1}\ldots\encTerm{M_n}})$,
and so we really have one of
$\invprove{\Gamma}{\Sigma}{(\encTerm{\app{c}{M_1\ldots M_n}})}{(\app{c}{M_1\ldots M_n})}{T}$
as desired.

\smallskip

\noindent{\it The term is of the form $\app{x}{M_1\ldots M_n}$ for some variable $x$.}
Then it must be that $\oftype{x}{\typedpi{y_1}{T_1}{\ldots\typedpi{y_n}{T_n}{T}}}\in\Gamma$.
From this we can be sure there are derivations of
$$\Gamma\vdash_{\Sigma}\oftype{M_i}{\subst{T_i}{M_1/y_1\ldots M_{i-1}/y_{i-1}}}$$
for each $i$.
So by the induction hypothesis 
$$\invprove{\Gamma}{\Sigma}{\encTerm{M_i}}{M_i}{\subst{T}{M_1/y_1\ldots M_{i-1}/y_{i-1}}}$$
is derivable for each $i$.
We can now compose these by \invvar~for a derivation of
$$\invprove{\Gamma}{\Sigma}{(\app{x}{\encTerm{M_1}\ldots\encTerm{M_n}})}{(\app{x}{M_1\ldots M_n})}{T}.$$
But $\encTerm{\app{x}{M_1\ldots M_n}}=(\app{x}{\encTerm{M_1}\ldots\encTerm{M_n}})$,
and so we really have one of
$\invprove{\Gamma}{\Sigma}{(\encTerm{\app{x}{M_1\ldots M_n}})}{(\app{x}{M_1\ldots M_n})}{T}$
as desired.
\end{proof}

We now have a method of re-encoding closed \hhf~terms back into LF,
which complements the term encoding in
Figure~\ref{fig:naive-translation}.
This reverse encoding provides the ability to run Twelf programs
containing such queries efficiently via Teyjus and then return the
results in a form which aligns with the specification formed in Twelf.

  \section{Towards Treating Existential Variables}
\label{sec:existential}

So far we have only considered the problem of finding inhabitants for
closed types.
In practice, Twelf also allows free variables to appear in types that
constitute queries.
These variables are considered to be existentially quantified in the
sense that answering the query requires finding substitutions for
these variables that make the type well-formed in
addition to providing an inhabitant for the resulting type.

Let us consider an example of the use of such variables based on the
signature for $append$ from Section~\ref{sec:lf}.
Suppose that we want to find the list $L$ that is the result of
appending $(cons~(s~z)~nil)$ to $(cons~z~nil)$.
We can have Twelf determine this list by posing the query
\[\oftype{M}{append~(cons~(s~z)~nil)~(cons~z~nil)~L}.\]
Now, Twelf must find a substitution for $L$ that yields a well-formed
type and simultaneously determine a value for $M$ that constitutes an
object of the resulting type.
To solve this query, Twelf will match it with the clause for
$appCons$, resulting in the new goal of constructing an inhabitant
$M'$ of the type $append~nil~(cons~z~nil)~L'$; this match will also
produce the binding $(cons~(s~z)~L')$ for $L$.
At this stage, Twelf will use the $appNil$  clause, resulting in a
solution to the overall goal with the binding $(cons~z~nil)$ for $L'$
and, therefore, of $(cons~(s~z)~(cons~z~nil))$ for $L$.
The inhabitant found for the original query will correspondingly be
\begin{tabbing}
\qquad\=$appCons~$\=\kill
\>$appCons~(s~z)~nil~(cons~z~nil)$\\
\>\>$(cons~z~nil)~(appNil~(cons~z~nil)).$
\end{tabbing}

The translation that we have described for LF types works without
modification also for types with free variables.
However, there is potential for the computational behavior of the
translated form to be different from that for Twelf because of the way
in which variables in types are instantiated.
Before we can discuss this in detail, it is important to
understand the way Twelf treats dependencies in types.
Let us assume the declarations introduced for representing natural
numbers and provide the following additions to the signature
\begin{tabbing}
\=\qquad\=\qquad\=\ \ \ \=\kill
\>\>$i$\>$:$\>$type$\\
\>\>$bar$\>$:$\>$nat\ra type$\\
\>\>$foo$\>$:$\>$\{X:i\}bar~z$
\end{tabbing}
Now we consider the query $\oftype{T}{bar~z}$.
Twelf will fail in this query under the following rationale:
The only way to construct an object of this type is by finding an
object of type $i$ and using this as an argument of $foo$.
However, there is no way to construct an object of type $i$.

An interesting thing happens if we change the definition of
$foo$ to make the argument variable actually appear in
the target type.
More specifically suppose that the definition of $foo$ is replaced
with the following:
\[\oftype{foo}{\{X:i\}bar~X}.\]
Now the query $\oftype{T}{bar~Y}$ returns the substitution
$$T=\oftype{[X:i]foo~X}{\{X:i\}bar~X}.$$
This solution can be interpreted as telling the user that if he/she
can provide something of type $i$, then there would be a term of the
required type.
From the specification we know that there cannot in fact be any terms
of type $i$, but Twelf does not try to find any terms of type $i$ and
instead leaves it as a constraint.

Twelf's behavior on dependent types can then be summed up as
follows.
If the argument variable dose not appear in the target type, thereby
signaling the absence of any real dependencies, proof search will
force the finding of an inhabitant for the corresponding type.
However, if the dependency is real, then the variable will be
instantiated only to the extent needed by other parts of the proof
search procedure; he actual finding of an inhabitant will not be
required for a successful solution.

Returning now to the comparison with the translated form, we see 
that the behaviors are convergent in the case where the argument
variable does not appear in the target type.
This is because the variable does not appear strictly in the target
type and hence the translation produces a \hhf~clause that forces the
search for an inhabitant.
We can see this, for example, by looking at the clause produced for
the first version of $foo$ which would be
$$\forall X(\hastype{X}{i}\supset\hastype{(foo~X)}{(bar~z)}).$$
When the variable does occur in the target type, however, we have two
different situations.
If the variable has at least one strict occurrence, then the clause
that is produced will not check for the type of the term instantiating
the variable and hence also will not force the search of an inhabitant
if the variable is uninstantiated.
This is seen, for example, from the clause resulting from the second
definition of $foo$, which will be
$$\forall X~\hastype{(foo~X)}{(bar~X)}.$$
However, if the variable does not have even one strict occurrence in
the type, then the translated version will force the search for an
inhabitant and the behavior corresponding to it will diverge from that
under Twelf.

We have conjectured that the computational behavior of Twelf on an LF
specification and of $\lambda$Prolog on a translated form of the LF
specification will be closely related if all the argument variables
have at least one strict occurrence in their types in the
specification.
In proving this conjecture, however, it is necessary also to take into
account unification behavior.
A complicating factor here is that unification in $\lambda$Prolog will
take place on ``type erased'' forms of terms in LF.
We believe, however, that this will not matter: the unifying
substitutions will be related also via a type erasure.
In particular we believe that the following claim holds.
\begin{claim}\label{claim:existential}
Suppose $M$ and $M'$ are terms of equivalent type.
Then $\sigma$ is a unifier for $M$ and $M'$ if and only if
$\sigma'=\encTerm{\sigma}$ is a unifier for $\encTerm{M}$ and
$\encTerm{M'}$.
\end{claim}
If we can show this claim, this will allow us to move between
unification of LF expressions and \hhf~terms freely because they will be
essentially equivalent.
Using this fact, we should be able to relate the operational semantics
of Twelf and \lprolog.
This would be done by recursively looking at each step and maintaining
a correspondence between the two.
From this we should be able to prove a strong form of correspondence
between the original LF specification satisfying the strictness
restriction and the translated version with respect to proof search
even in the presence of existential variables.

When the restriction is not satisfied, the behaviors diverge as we
have noted earlier.
To understand the nature of the divergence, let us consider the
following example signature.
\begin{tabbing}
\=\qquad\=\qquad\=\ \ \ \=\kill
\>\>$nat$\>$:$\>$type.$\\
\>\>$z$\>$:$\>$nat.$\\
\>\>$s$\>$:$\>$nat\ra nat.$\\
\>\>$bar$\>$:$\>$nat\ra type $\\
\>\>$foo$\>$:$\>$\{Y:nat\}\{F:nat\ra nat\}bar~(F~Y)$
\end{tabbing}
Consider the query $\oftype{T}{bar~z}$.
Twelf is unable to resolve the equation $z=F~X$ and so cannot supply a
solution for this query.
However, under the translation we can determine the obvious solutions
in a systematic way.
The clause for $foo$ under the translation would have the form
\begin{tabbing}
\=\qquad\=\qquad\=\kill
\>\>$\forall Y(\hastype{Y}{nat}\supset$\\
\>\>$\forall F(\forall X1(\hastype{X1}{nat}\supset\hastype{(F~X1)}{nat})\supset$\\
\>\>\>$\hastype{(foo~F~Y)}{(bar~(F~Y))}))$
\end{tabbing}
and the query is $\hastype{T}{(bar~z)}$.
Backchaining on the clause for $hastype$ to solve this query still
requires us to unify $z$ and $F~Y$.
However, this time we have a method for finding valid
substitutions to consider for $F$ and $Y$ in the course of solving the
unification problem.
First a term of type $nat$ must be determined for $Y$ using a query
such as $\hastype{Y}{nat}$ and once fixed, a substitution for $F$ will
be formed using 
$$\forall X_1(\hastype{X_1}{nat}\supset\hastype{(F~X_1)}{nat}).$$
If these substitutions satisfy $F~Y=z$ they can be considered a valid solution.
Working through a few examples, if we first fix $Y=z$ by matching with
the generated clause for $z$ in the translated signature, both
$\lambdax{x}{x}$ and $\lambdax{x}{z}$ will be substitutions for $F$
for which $(F~Y)=z$.
Next we would fix $Y=(s~z)$ as a substitution, and so $\lambdax{x}{x}$
no longer satisfies our constraint.
Thus there is only one substitution for $F$ which is valid, $F=\lambdax{x}{z}$.
In fact, for every successive natural number only $F=x\ z$ will
satisfy $F~Y=z$.
In this systematic manner we are able to generate the possible
substitutions using the \lprolog~program.

  \section{Conclusion}
\label{sec:conclusion}

This paper continues the work in~\cite{snow10ppdp} on
translating LF specifications into \hhf~formulas.
We have extended that work by defining a richer notion of strictness
that can potentially lead to the elimination of more type checking
and can thereby be the basis for a more efficient and clearer
translation.
We have also presented a procedure for translating \hhf~terms back to
LF objects in the context of the original LF specifications.
This inverse translation provides a means for taking terms
generated by running the translated specification and constructing the
corresponding LF term.
Finally, we have analyzed the situation where existential variables
appear in LF types; this situation leads to type reconstruction in the
Twelf system.
Although we have still to formalize our observations, our analysis
suggests that when the types in an LF specification are such that
every dependently quantified variable appears strictly in the target
type, then the behavior of $\lambda$Prolog over the \hhf~ translation
will mimic that of Twelf over the LF specification.

In future work, we would like to translate the results of this paper
into practical applications.
We have already modified the code in the Parinati system
\cite{parinati.website} to take into account the extended strictness
check described in Section~\ref{sec:refined}.
We are currently in the process of testing the resulting system and 
assessing whether the new cases it covers are ones that have
real efficiency benefits.
In a related direction, we would like to use our ideas to provide an
alternative implementation of the logic programming part of the Twelf
system.
This new system would take LF specifications and queries, translate
them into their $\lambda$Prolog counterparts and then use the inverse
translation to return the results in a form understandable in the LF
setting.

Another direction for continued work concerns the treatment of
existential variables in types.
One task is to formalize the observations made in
Section~\ref{sec:existential}.
This would involve, for example, providing a proof of
Claim~\ref{claim:existential}. 
In Section~\ref{sec:existential} we also noted that when existential
variables are present there are situations in which the behavior under
the translation would be different from that under the original LF
specification.
We would like to understand these situations better: we feel that the
translated form might actually give us better control over
computational behavior than the LF version in these cases and would
like to substantiate this aspect if it is actually true.

The work in this paper has dealt exclusively with reasoning from the
specifications written in LF.
A completely different direction to pursue is that of reasoning {\em about}
LF specifications.
To understand the difference, we might consider the specification of
natural numbers and the plus relation that was provided in
Section~\ref{sec:lf}.
Based on these specifications, we might want to show something about
the plus relation.
For example, we may want to show that given two natural numbers $m$
and $n$ there is always another natural number $k$ that is $m$ plus
$n$.
This clearly cannot be shown by solving any query from the
specification.
Rather, it involves proving something about {\em all}
queries that can be made against the specification.
The Twelf system allows such reasoning to be realized through tools
for showing the totality of specifications written in it.
For example, the property in question about $plus$ can be established
by showing that for any $M$ and $N$ there is always a value $K$ for
which the goal $plus\ M\ N\ K$ will succeed.
Totality checking does not provide an explicit proof of the property
since there is no explicit logic supporting this style of reasoning.
The translation from LF to the \hhf~logic suggests an alternative
path: we can think of also translating totality checking into an
explicit proof in the Abella system~\cite{abella.website} that
supports the capability of logic based reasoning about \hhf~
specifications.
As a continuation of this work, we would like to explore the extension
of the translation to this kind of meta-theoretic reasoning.
Another direction that is much more challenging is to see if the
Abella logic can be used to directly reason over LF-style
specifications rather than having to do this via translations.

\acks
This work has been partially supported by the NSF Grant CCF-0917140.
Opinions, findings, and conclusions or recommendations expressed in this paper
are those of the authors and do not necessarily reflect the views of the
National Science Foundation.

  \bibliographystyle{abbrvnat}
  \bibliography{references/master}

\begin{thebibliography}{9}
\providecommand{\natexlab}[1]{#1}
\providecommand{\url}[1]{\texttt{#1}}
\expandafter\ifx\csname urlstyle\endcsname\relax
  \providecommand{\doi}[1]{doi: #1}\else
  \providecommand{\doi}{doi: \begingroup \urlstyle{rm}\Url}\fi

\bibitem[Church(1940)]{church40}
A.~Church.
\newblock A formulation of the simple theory of types.
\newblock \emph{J. of Symbolic Logic}, 5:\penalty0 56--68, 1940.

\bibitem[Felty and Miller(1990)]{felty90cade}
A.~Felty and D.~Miller.
\newblock Encoding a dependent-type $\lambda$-calculus in a logic programming
  language.
\newblock In M.~Stickel, editor, \emph{Proceedings of the 1990 Conference on
  Automated Deduction}, volume 449 of \emph{LNAI}, pages 221--235. Springer,
  1990.

\bibitem[Gacek(2009)]{abella.website}
A.~Gacek.
\newblock The {A}bella system and homepage.
\newblock \url{http://abella.cs.umn.edu/}, 2009.

\bibitem[Harper et~al.(1993)Harper, Honsell, and Plotkin]{harper93jacm}
R.~Harper, F.~Honsell, and G.~Plotkin.
\newblock A framework for defining logics.
\newblock \emph{Journal of the ACM}, 40\penalty0 (1):\penalty0 143--184, 1993.

\bibitem[Miller and Nadathur(2012)]{miller12proghol}
D.~Miller and G.~Nadathur.
\newblock \emph{Programming with Higher-Order Logic}.
\newblock Cambridge University Press, June 2012.
\newblock \doi{10.1017/CBO9781139021326}.

\bibitem[Nadathur and Miller(1988)]{nadathur88iclp}
G.~Nadathur and D.~Miller.
\newblock An {Overview} of {$\lambda$Prolog}.
\newblock In \emph{{Fifth International Logic Programming Conference}}, pages
  810--827, Seattle, Aug. 1988. MIT Press.

\bibitem[Qi et~al.(2008)Qi, Gacek, Holte, Nadathur, and Snow]{teyjus.website}
X.~Qi, A.~Gacek, S.~Holte, G.~Nadathur, and Z.~Snow.
\newblock The {T}eyjus system -- version 2, Mar. 2008.
\newblock http://teyjus.cs.umn.edu/.

\bibitem[Snow(2010)]{parinati.website}
Z.~Snow.
\newblock {P}arinati.
\newblock \url{http://www.cs.umn.edu/~snow/parinati}, 2010.

\bibitem[Snow et~al.(2010)Snow, Baelde, and Nadathur]{snow10ppdp}
Z.~Snow, D.~Baelde, and G.~Nadathur.
\newblock A meta-programming approach to realizing dependently typed logic
  programming.
\newblock In \emph{ACM SIGPLAN Conference on Principles and Practice of
  Declarative Programming (PPDP)}, pages 187--198, 2010.

\end{thebibliography}
\end{document}